\documentclass[superscriptaddress,prb,super=false,
showpacs,floatfix,fleqn,longbibliography]{revtex4-1}

\UseRawInputEncoding
\setcitestyle{numbers,square}

\usepackage[dvips]{graphicx}
\usepackage{epsfig} 
\usepackage{subfig}
\usepackage{color}
\usepackage[normalem]{ulem}
\usepackage{amsmath,amsfonts,amssymb,bm}
\setcitestyle{numbers,square}
\UseRawInputEncoding

\begin{document}
\title{Super diffusive length dependent  thermal conductivity in  one-dimensional materials  with structural defects: longitudinal to transverse phonon scattering leads to  $\kappa \propto L^{1/3}$ law. }
\author{Alexander L. Burin} 
\email[]{aburin@tulane.edu}
\affiliation{Department of Chemistry, Tulane University, New
Orleans, LA 70118, USA}
\date{\today}
\begin{abstract}
Structural defects in one-dimensional heat conductors couple longitudinal (stretching) and transverse (bending)  vibrations.  This coupling results in the scattering of longitudinal phonons to transverse phonons and backwards.  We show that the decay rate  of longitudinal phonons due to this scattering scales with their frequencies as $\omega^{3/2}$ within the long wavelength limit ($\omega \rightarrow 0$), which is more efficient scattering  compared to the  traditionally considered Rayleigh scattering within the longitudinal band  ($\omega^2$). This scattering results in temperature independent  thermal conductivity depending on the size as   $\kappa  \propto L^{1/3}$ for sufficiently long materials. This  predicted length dependence  is observed in nanowires, though the temperature dependence is seen there  possibly because of deviations from pure one-dimensional behavior.   The significant effect of interaction of longitudinal phonons with transverse phonons  is  consistent with the earlier observations of a substantial suppression of thermal energy transport by kinks, obviously leading to such interaction,  though anharmonic interaction can also be significant.   
\end{abstract}

\maketitle

\section{Introduction}
\label{sec:Intr}


Although polymer materials have a low thermal conductivity and   are used as thermal insulators,   the thermal conductivity of aligned polymer molecules is very  high in the alignment direction and can be comparable or even bigger than that of metals due to covalent bonds \cite{1977ChoyReviewPolymerThCond,2010PolyethNanoFibHigThCond,2022QuantPhTrDvira}.   Long polymer molecules used as heat conductors  are promising candidates for  heat control  management in micro/nanodevices   \cite{NitzanScience03,AbeScience07,SegalNitzan03}.   Other one-dimensional heat conductors including nanowires  \cite{2018SiNanowLdepThCond,2021SuperDiffN033}  and quasi-one-dimensional Van der Waals materials \cite{VanDerWaalsMaterReview2022,20221DPhTaSe,2023DopingEnhPhTransp,2025VanDerWaalsAtChainRevChin}  are also of great interest  for heat control applications.   Yet, in contrast to electronic transport , our understanding and, consequently,  ability to control this transport is rather limited, particularly, because of the lack of proper theoretical understanding of actual mechanisms of the propagation of energy carriers, i. e. phonons, in low dimensional materials.  Particularly, the origin of the superdiffusive phonon transport there is not well understood yet and we hope this work sheds some light onto this long-standing problem. 

Numerous measurements of thermal conductivity in one-dimensional materials   suggest that superdiffusive transport often takes place there. This transport is  characterized by the thermal conductivity $\kappa$ growing with system sizes $L$ as 
\begin{eqnarray}
\kappa \propto L^{\alpha}
\label{eq:NonFourBehav}
\end{eqnarray}
with the exponent $\alpha$ varying from $0.12$ to $1$ \cite{1998QuantThermCond,1998HeatTranspRev,Turbulence02,1998HeatTranspRev,2015SuperHeatDiffMomCons,2017BallTr4KSiNanow033,2018SiNanowLdepThCond,2021SuperDiffN033,2024prlnanopartonpolymsuperdiff} (see also  recent reviews \cite{2003RevClassLowD,2021RevLiviPedagog,2021ballist,2022QuantPhTrDvira}  and references therein).      The largest exponent $\alpha =1$  is realized for the ballistic propagation of all contributing  phonons,    where the thermal conductivity is determined by the quantum of thermal conductance  \cite{1998QuantThermCond,1998HeatTranspRev,2000NatQuantThermCond}.     This ballistic regime also  takes  place in organic polymers for optical phonons even at room temperature because of the weakness of anharmonic interactions compared to  harmonic ones  \cite{ab15ballistictranspexp,ab19IgorReview}.

These superdiffusive behaviors are consistent with the theoretical analysis of longitudinal phonon transport in the atomic chains  characterized by  the Fermi-Pasta-Ulam-Tsingou (FPUT) model used in the seminal  numerical experiment \cite{FPUclassic,2004TrVibrMD,2013FPUChainExactMendl,2013FPUTAnalNum}.  Thermal conductivity increases with the system length $L$ due to the fraction of ballistic phonons with longest wavelengths  propagating without scattering.    Such phonons exists as Goldstone modes  because of the symmetry of the system with respect to an identical infinitesimal  simultaneous translations of all participating atoms. Their scattering rate disappears in the infinite  wavelength limit corresponding to  a zero phonon frequency, where the displacement of nearby atoms become nearly  identical.

At low temperature, where anharmonic interaction is negligible, the superdiffusive behavior emerges due to scattering by defects. Particularly, the length dependence  $\kappa \propto \sqrt{L}$ is expected  due to  Rayleigh scattering of longitudinal phonons  caused by structural and mass defects  \cite{Lebowitz67,LEBOWITZ00,20011DGlass} leading to the phonon decay rate  $\gamma$ scaling with its frequency as $\gamma \propto \omega^{2}$.    The frequency dependence   $\gamma(\omega) \propto \omega^{2}$  corresponds to  the generalized Rayleigh scattering in $d$-dimensions $\gamma(\omega) \propto \omega^{d+1}$ leading to the well-known  $\omega^4$ behavior  in three dimensions.  Since a speed of longitudinal sound $c_{l}$ is constant in the long wavelength limit,  the phonon mean free path scales as $l = c_{l}/\gamma  \propto \omega^{-2}$ if the frequency $\omega$ tends to $0$.  Consequently, the phonon  mean free path exceeds the heat conductor  length ($l>L$) at very low frequencies $\omega < \omega_{L}$, where $\omega_{L} \propto L^{-1/2}$.  In this regime,  the  thermal conductivity between the left and right leads having the temperatures $T_{R}$ and $T_{L}$, respectively, and connected by one-dimensional conductor of the length $L$  is defined using the generalization of the Landauer formula  \cite{Landauer57Classic} for the thermal conductivity \cite{1988EquatForThermCond,1998QuantThermCond,1998HeatTranspRev} 
\begin{eqnarray}
\kappa = \frac{L}{2\pi (T_{R}-T_{L})}\int \mathcal{T}(\omega)(n_{R}(\omega)-n_{L}(\omega))\hbar\omega d\omega, 
\label{eq:kappaDvira}
\end{eqnarray}  
where $\mathcal{T}(\omega)$ is a frequency dependent phonon transmission between the left and right leads,  the notations $n_{R,L}(\omega)$ stand for the phonon population numbers $n_{L,R}=1/(e^{\frac{\hbar\omega}{k_{B}T_{R,L}}}-1)$  and the integration over frequency $\omega$ results from the integration over the wave vector after the substitution $d\omega=vdq$, where $v$ is the phonon propagation velocity  \cite{SegalNitzan03,TransportTextbird2002transport}.   
Assuming small temperature difference  $|T_{L}-T_{R}|\ll T_{R}$ and sufficiently long  conductor, so that  $\hbar\omega_{L} < k_{B}T_{R,L}$, we estimate the thermal conductivity given by Eq. (\ref{eq:kappaDvira}) using only the contribution of purely ballistic phonons with low frequencies $\omega < \omega_{L}$ as 
\begin{eqnarray}
\kappa = \frac{k_{B}}{2\pi}\omega_{L}L\propto L^{1/2}, 
\label{eq:ThermCondAnom}
\end{eqnarray} 

Further investigation of phonon transport in classical anharmonic FPUT-like models has led to predictions of different exponents $\alpha$ in the  thermal conductivity length dependence including $\alpha =2/5$ obtained by  applying the mode-coupling theory \cite{1998LepriAnomalThCondFPUCl} or  $\alpha =1/3$ that results from  hydrodynamic fluctuations caused by the momentum conservation law \cite{Turbulence02} or anharmonic interaction with transverse vibrations  \cite{2004TrVibrMD}  lacking in the FPUT model but included in the molecular dynamic simulations.    

Thermal conductivity of one-dimensional polymer chains was considered  using extensive molecular dynamic simulations  \cite{2004TrVibrMD,2020MDLangEnergyTrNitzan,2022highthcond-diam-nanow,2024SubSuperDiffTranspThRectif} (for more detail see the recent review \cite{2023MDSimulRev}  and references therein).   These simulations show the features not seen in the FPUT-like models including dramatic sensitivity to defects, especially including kinks \cite{ThermCondKinks2019Xuhui}.  Particularly,  according to those simulations kinks result in the transport, obeying the Fourier law (similar outcome was found for defects like breaks \cite{2023ChainWithBrks}).    The comprehensive molecular dynamics investigation of carbon nanotubes \cite{2021prlCarmonNanotubeSimulHighConvThermCond} has led to the conclusion that thermal conductivity behaves according to the Fourier law for sufficiently long tube lengths.   

Thus, there is a difference between the analytical results obtained using the simplified FPUT-like model and the molecular dynamics approach attempting to treat materials as they are.  In our opinion, this difference is caused by the restriction of the FPUT model  to only longitudinal vibrations, while the molecular dynamics methods treat all significant vibrational modes.  At low temperature, where the thermal energy $k_{B}T$ is much smaller than the maximum quantum energy $\hbar\omega_{\rm l,max}$ ($\omega_{\rm l,max}$ is the maximum frequency of a longitudinal phonon expressing the width of the vibrational band)  the consideration is restricted to longitudinal phonons. This is because these phonons have no bandgap and there always exists a substantial fraction of such phonons possessing energy $\hbar\omega$ comparable or less than the thermal energy so they are capable of efficiently transferring heat.  However,  there are three other gapless acoustic phonon bands corresponding to torsional and transverse vibrations  \cite{landau1986theory,PolimVibr1994Book,2004TrVibrMD,Chico2006VibrCarbNanot,2010TransvVibr1D}.  The torsional 
modes possess a sound like spectrum $\omega_{\rm tors}(q)=c_{\rm tors}q$ with the speed of sound $c_{\rm tors}$ usually being smaller compared to that of the longitudinal sound  $c_{l}$ (here $q$ stands for the phonon wavevector.)  Two transverse acoustic bands enumerated by the index $\mu=1,2$ possess the quadratic spectra $\omega_{\mu}(q)=c_{\mu}aq^2$ where $a$ is a lattice period and the velocity parameter $c_{\mu}$ estimates the maximum propagation velocity of transverse phonons realized at $q \sim 1/a$.   Torsional and transverse phonons are slower compared to longitudinal phonons so at the first glance they can be neglected.

Although the longitudinal phonons are the most  efficient heat carriers, the scattering of longitudinal phonons by slower torsional or transverse phonons can be much stronger than their scattering by themselves, cf. Refs.  \cite{ab20Transv,ab2023Cherenkov,ab242stagedec}.  
This is because torsional and transverse phonons   possess  larger densities of states compared to longitudinal phonons.  Particularly for  transverse phonons their density of states diverges in the long-wavelength limit  as  $\omega^{-1/2}$.   Therefore, we expect that longitudinal phonon transport in one-dimensional heat conductors  can be  dramatically sensitive to  other acoustic phonon bands.  Here, we demonstrate  that  at low temperatures where anharmonic interactions can be approximately neglected  the dominating phonon  scattering mechanism turns out to be  the defect induced scattering of  longitudinal phonons to transverse phonons.   Below we show that for longitudinal phonons, this scattering results in the frequency dependent phonon decay rate depending on the frequency within the long wavelength limit as   
\begin{eqnarray}
\gamma(\omega) \propto \omega^{3/2}.  
\label{eq:NewLaw}
\end{eqnarray}
For this specific dependence  the phonons with frequencies $\omega<\omega_{L}\propto L^{-2/3}$ propagate ballistically,  contributing to the thermal conductivity length dependence as $\kappa \approx k_{B}\omega_{L}L \propto L^{1/3}$  (cf. Eq. (\ref{eq:ThermCondAnom}).  This dependence is similar  to the earlier predictions \cite{Turbulence02,2004TrVibrMD}, but it emerges  due to scattering by defects, dominating in the low temperature limit.  We do expect that anharmonic interaction of longitudinal phonons with transverse phonons will be significant at higher temperatures and can be responsible for the behaviors observed in molecular dynamics simulations, but leave this consideration for the future.

The paper is organized as following.   In Sec.  \ref{sec:LongWvlPhen} we derive the main result of the present work, Eq. (\ref{eq:NewLaw}),  using the phenomenological model of interaction of longitudinal and transverse modes in the form $\beta(\partial u_{x}/\partial x)\times (\partial^{2}u_{y}/\partial x^2)$ where $x$ is the direction of the molecular axis and displacements $u_{x}$ and $u_{y}$ stand for longitudinal and transverse displacements, respectively.  This coupling is valid within the long wavelength limit.  We evaluate the phonon decay rate using  the Fermi Golden rule similarly to Refs.  \cite{ab2023Cherenkov,ab242stagedec} where this approach is justified by the numerical simulations of phonon decay.   In Sec.  \ref{sec:Fence}, we show  the presence of coupling of transverse and longitudinal modes within  the  toy ``fence'' model with defects, numerically  evaluating the reflection  coefficient by a single defect.  We show that the reflection coefficient possesses the frequency dependence in the form of Eq.   (\ref{eq:NewLaw}).  In Sec.  \ref{sec:Exp} we briefly discuss the relevance of the proposed theory to existing experimental data.  The results are summarized in the brief conclusions formulated in Sec. \ref{sec:Concl}.


\section{Scattering of longitudinal phonons to transverse phonons and backwards within the long-wavelength limit.}
\label{sec:LongWvlPhen}

\subsection{Model}
\label{sub:LonWvModel}

Here, we consider a one-dimensional heat conductor   at sufficiently  low temperature, where anharmonic interactions scattering can be approximately discarded  and the phonon transport is determined by phonon  scattering by defects emerging within the harmonic approximation.    In addition to scattering,  third and fourth order anharmonic interactions of longitudinal phonons  can affect the phonon spectrum since they contain resonant terms.  Indeed  for linear phonon spectrum $\omega(q) \approx c_{l}q$ some scattering processes conserving wavevector  also approximately conserve energy, i. e.  $\omega(q_{1}) + \omega(q_{2})\approx \omega(q_{1}+q_{2})$ for identical signs of $q_{1}$ and $q_{2}$ so the third and fourth order anharmonic interaction lead to resonant scattering.     However, these resonant terms result in corrections of order of $\omega^2$ and $\omega^3$ \cite{SchickFPULuttLiq68} to the free particle spectra, which  emerges after fermionization.   Consequently, we  ignore these effects when comparing  to the phonon decay rate scaling as $\omega^{3/2}$ in the long wavelength limit $\omega \rightarrow 0$.   Therefore, our consideration of the pure harmonic model is well justified at sufficiently low temperatures, where  anharmonic scattering  is  negligible.  

For the sake of simplicity we restrict our consideration to longitudinal displacements   along the heat conductor  axis $x$ and transverse displacements in one perpendicular direction $y$.  The omitted displacements in the  $z$ direction will add another transverse phonon band.    This band does  not modify our results for the thermal conductivity of longitudinal phonons  qualitatively, because it adds a similar scattering channel  (to the other transverse band),  characterized by the same frequency dependence. 

A consideration of pure one-dimensional transport is well justified  in isolated molecular chains.   To our knowledge thermal conductivity length dependence has been  probed only in polymer-grafted nanoparticle melts \cite{2024prlnanopartonpolymsuperdiff}, while many other measurements were carried out in nanowires \cite{2018SiNanowLdepThCond,2019IncrThCondSiGe,2021SuperDiffN033,2024IncrThCondSiCarbide} having a diameter $d$  of at least a few nanometers.  These nanowires  can be treated as one-dimensional only if the phonons with non-zero  wavevectors perpendicular to the nanowire axis are frozen out.  This requires the minimum energy of phonons propagating in the  transverse direction to exceed the thermal energy,   i. e.  
\begin{eqnarray}
k_{B}T <  \hbar \frac{c_{\rm l}}{d}. 
\label{eq:AreaConstr}
\end{eqnarray}

As discussed in the introduction, we expect that the thermal conductivity is determined by the low frequency phonons  possessing the long wavelengths $\lambda \gg a$, where $a$ is the lattice period, and, consequently, small wavevectors $q=2\pi/\lambda  \ll \pi/a$, where $\pi/a$ is the maximum wavevector.   Then we can consider  only displacements $u_{\mu}$ with small wavevectors redefining them in the integral form as \cite{StatPhysLandau} 
\begin{equation}
\widehat{\tilde{u}_{\mu}}(x) = \frac{a}{2\pi}\int_{-q_{0}}^{q_{0}} u_{\mu q}e^{-iqx},  ~ \mu=x, ~ y,  ~  u_{\mu q}=\frac{1}{a}\int dxu_{\mu}(x) e^{iqx}.
\label{eq:DisplRen}
\end{equation}
The same procedure is assumed for the momenta corresponding to the displacements under consideration.  

 Within the harmonic approach the Hamiltonian is expanded  with respect to squared displacements (there is no linear terms because zero displacements should correspond to the energy minimum).  Since identical displacements of all atoms $u_{\mu}=u_{0}$ does not modify the system's energy,  the Hamiltonian can depend only on displacement derivatives with respect to the molecular axis direction $x$,  such as  $\partial^{m} u_{\mu}/\partial x^m$,  $m\geq 1$. Within the long wavelength limit we should leave the  terms containing  the smallest number of derivatives $m$  \cite{landau1986Elasticitytheory}.   For longitudinal displacements  $u_{x}$ the only $m=1$ derivative responsible for the compressive strain should be left. This yields  the longitudinal vibration  Hamiltonian in the form \cite{StatPhysLandau} (in the absence of defects) 
 \begin{eqnarray}
 \widehat{H}_{\rm l}= \frac{1}{2a}\int dx \left(\frac{p_x^2}{M} + A_{\rm l}\left(\frac{\partial u_{x}}{\partial x}\right)^2\right). 
 \label{eq:LongH}
 \end{eqnarray}
Here $A_{\rm l}$ is the longitudinal vibration force constant,  $M$ stands for the mass of elementary cell  and we discard  ``tilde'' notation used in  Eq. (\ref{eq:DisplRen}) that is applied to  all coordinates and momenta.  

The uniform  transverse displacement $u_{y} \propto x$ does not modify the system's energy in the second order in $u_{y}$ since it acts like the rotation of the whole system  \cite{landau1986Elasticitytheory}.  Consequently,  the energy of  bending, which emerges due to transverse displacements,  is determined by the squared second derivative of the displacement $u_{y}$ expressing a squared curvature of the molecular axis and the Hamiltonian of transverse vibrations takes the form \cite{landau1986Elasticitytheory} 
 \begin{eqnarray}
 \widehat{H}_{\rm tr}= \frac{1}{2a}\int dx \left(\frac{p_y^2}{M} + A_{\rm tr}a^2\left(\frac{\partial^2 u_{y}}{\partial y^2}\right)^2\right),
 \label{eq:TrH}
 \end{eqnarray}
and the parameter $A_{\rm tr}$ expresses the  force constant for transverse vibrations. 

Structural defects like kinks mix up longitudinal and transverse modes.  The simplest form of their interaction allowed by the rotational and translational invariance  is given by the product $(\partial u_{x}/\partial x)\times (\partial^{2} u_{y}/\partial x^2)$.  This interaction   violates   both reflection symmetry  ($y\rightarrow -y$) and inversion symmetry ($x,y\rightarrow -x, -y$) so the system with defects should not possess these symmetries.  The example of such a defect leading to the desirable interaction is given in  Sec. \ref{sec:Fence}  within the toy ``fence'' model of the molecule.  

We introduce the defect Hamiltonian in the form 
  \begin{eqnarray}
 \widehat{V}_{\rm l,tr}= Ba\sum_{i} \left.\frac{\partial u_{x}}{\partial x}\frac{\partial^{2} u_{y}}{\partial x^2}\right\rvert_{x=x_i}, 
 \label{eq:DefHltr}
 \end{eqnarray}
where $B$ is the force constant characterizing interaction induced by  defects. We assume that they are  located randomly at positions $x_{i}$ with the number density $n$.  Also,  for the sake of simplicity we assume that they are identical.   

The interaction in Eq.  (\ref{eq:DefHltr}) can be originated only from structural defects.  Mass defects, e. g. due to isotopes, do not result in the direct overlap of transverse and longitudinal bands since  kinetic energy does not contain   products of different projections of momentum in longitudinal and transverse directions.  Therefore,  we expect  that mass defects result mostly  in the Rayleigh scattering leading to the thermal conductivity size dependence $\kappa \propto N^{1/2}$. 

Defects also lead to scattering of phonons inside each band.  This scattering is characterized by the Hamiltonian 
  \begin{eqnarray}
 \widehat{V}_{\rm def}= \left.D\sum_{i} \left(\frac{\partial u_{x}}{\partial x}\right)^2\right\rvert_{x=x_i}+\left.Fa^2\sum_{i} \left(\frac{\partial^{2} u_{y}}{\partial x^2}\right)^2\right\rvert_{x=x_i}. 
 \label{eq:DefHins}
 \end{eqnarray}

Before the further consideration, this remark is in order.  The model formulated above does not include short wavelength phonons and anharmonic interactions between them.  We assume that these interactions are included in  the definitions of renormalized force constants $A_{\rm l, tr}$, $B$, $D$ and $E$. Obviously, these interactions affect both spectra of transverse and longitudinal phonons and their coupling induced by defects.  For instance,  heavy mass   defects lead to the low frequency resonance due to the quasi-local modes \cite{KaganIoselevskiiQuasiLoc,1965QuasilocSignofNextNeighb,1988IYPVibrLoc8,ab90JETPLettPhonons,abiypphysrep,2019AnalyticsForRandMass,2023QuasilocModIsotopDefect}. In that case, the force constants such as  $D$ and $F$ in Eq. (\ref{eq:DefHins}) are substantially renormalized compared to their input values,  while the form of the interaction remains the same. 
We assume that the renormalization of the Hamiltonian due to the anharmonic interactions with high energy phonons  modifies all force constants in Eqs.  (\ref{eq:DefHltr}), (\ref{eq:DefHins}),  but does not change the form of the Hamiltonian. 
 
It is convenient to express  the vibrational  Hamiltonian  in terms of  creation and annihilation operators of longitudinal and transverse phonons   ($b_{q}^{\dagger},  b_{q},  c_{q}^{\dagger},  c_{q}$, respectively) using the wavevector $q$ representation.  In this representation the full Hamiltonian can be expressed as 
\begin{eqnarray}
\widehat{H} = \hbar \sum_{q}^{|q|<q_{0}}\left( c_{l}q b_{q}^{\dagger}b_{q} + c_{\rm tr}aq^2  c_{q}^{\dagger}c_{q}\right)+
\frac{\hbar}{N} \sum_{q}^{|q|<q_{0}} \sum_{q'}^{|q'|<q_{0}}\sum_{i} iba^{3/2}|q'|\sqrt{|q|}\left(b_{q}^{\dagger}c_{q'}-b_{-q}c_{-q'}^{\dagger}\right)e^{ix_{i}(q-q')}+
\nonumber\\
+\frac{\hbar}{N}\sum_{q}^{|q|<q_{0}} \sum_{q'}^{|q'|<q_{0}}\sum_{i} \left(da\sqrt{|qq'|}b_{q}^{\dagger}b_{q'}+ea^2|qq'|c_{q}^{\dagger}c_{q'}\right)e^{ix_{i}(q-q')},
\nonumber\\
c_{\rm l}=\sqrt{\frac{A_{\rm l}}{M}}, ~ c_{\rm tr}=\sqrt{\frac{A_{\rm tr}}{M}},  ~ b=\frac{B}{2M\sqrt{c_{\rm l}c_{\rm tr}}a},  ~ d=\frac{D}{Mc_{\rm l}a}, ~ f=\frac{F}{Mc_{\rm tr}a}.
\label{eq:FullH}
\end{eqnarray}
All redefined interaction constants $b$,  $d$ and $f$ are expressed in the frequency units.  In Eq.  (\ref{eq:FullH}) only resonant terms containing simultaneous creation and annihilation of phonons  are left  since only those terms can lead to the energy conserving scattering.  

We do not consider the possible existence of the  coupling of longitudinal and transverse phonons not related to the defects  because it cannot result in the phonon scattering.  Indeed,  this coupling will add the terms having the form  $q^3b_{q}^{\dagger}c_{q}$ or $q^3c_{q}^{\dagger}b_{q}$ to the Hamiltonian Eq.  (\ref{eq:FullH}).  The scattering induced by such terms modifies the system energy by the difference of phonon energies $c_{l}q - c_{\rm tr}aq^2$, which is much greater than their coupling strength scaling as $q^3$ within the long wavelength limit. 

Scattering by defects does not conserve wavevectors and therefore can conserve energy,   as   shown below in Sec. \ref{sub:FGRmod}, where we estimate the phonon decay rate due to the scattering by defects  using the Fermi Golden rule. 

\subsection{Calculations  of phonon decay rate using Fermi Golden rule}
\label{sub:FGRmod}

Here, we calculate frequency dependent  decay rates $\gamma_{\rm l,tr}$ of longitudinal and transverse phonons, respectively,  using the Fermi Golden rule.  We assume that the maximum cutoff wavevector $q_{0}$ is small enough so  that  the first Born's approximation expressed by the Fermi golden rule gives a sufficiently accurate estimate of the phonon lifetime.


We begin with the calculations of the decay rate of a longitudinal phonon with the frequency $\omega$ (wavevector $q=\omega/c_{l}$) due to  its coupling with the transverse phonons. This coupling is  determined by the interaction in Eq (\ref{eq:DefHltr}) and expressed by the third term of the full Hamiltonian in Eq. (\ref{eq:FullH}).  The scattering by each defect can lead to the transition of this phonon to the transverse phonon with approximately the same frequency (due to the energy conservation) and the wavevector $q'= \pm\sqrt{\omega/(c_{\rm tr}a)}$.  Consider the backwards scattering   corresponding to the negative wavevector $q'$.  The probability of the forward scattering is identical. 

The rate of backwards scattering  by a single defect is given by the Fermi Golden rule in the form 
\begin{eqnarray}
W_{\rm 1}=\frac{qa^4b^2}{N}\int_{-q_{0}}^{0}dq'\delta(\omega-c_{\rm tr}aq'^2)q'^2= \omega^{3/2}\frac{b^2}{2\Omega_{\rm tr}^{3/2}\Omega_{\rm l}N}, ~ \Omega_{\rm l,tr} =\frac{c_{\rm l,tr}}{a},
\label{eq:SingleDef}
\end{eqnarray}
where the frequency $ \Omega_{\rm l,tr}$ is of order of the longitudinal or transverse phonon bandwidths, respectively.  

By adding all defect contributions and also both backwards and forward scattering, we come up with the final expression for the longitudinal phonon decay rate in the form (remember, that $n$ is the number density of defects per a unit length) 
\begin{eqnarray}
\gamma_{\rm l,tr} =  na\frac{b^2}{\Omega_{\rm l}}\frac{\omega^{3/2}}{\Omega_{\rm tr}^{3/2}}.  
\label{eq:LTRDec}
\end{eqnarray}

Eq.  (\ref{eq:LTRDec}) is the main result of the present work that leads to the new thermal conductivity behavior.   Before considering the thermal conductivity, we evaluate other contributions to the phonon lifetimes using the Fermi Golden rule similarly to Eq. (\ref{eq:SingleDef}). Since the calculations are quite similar to the case of longitudinal to transverse phonon scattering, we simply  report the final results for three other remaining lifetimes $\gamma_{\rm l,l}$,   $\gamma_{\rm tr,l}$ and  $\gamma_{\rm tr,tr}$ for longitudinal to longitudinal, transverse to longitudinal and transverse to transverse phonon scattering rates. These rates are defined as 
 \begin{eqnarray}
\gamma_{\rm l,l} = 2na \frac{d^2}{\Omega_{\rm l}}\frac{\omega^{2}}{\Omega_{\rm l}^{2}}, ~ \gamma_{\rm tr,l} = 2na\frac{b^2}{\Omega_{\rm tr}}\frac{\omega^{2}}{\Omega_{\rm l}^{2}}, ~  \gamma_{\rm tr,tr} =na\frac{f^2}{\Omega_{\rm tr}}\frac{\omega^{3/2}}{\Omega_{\rm tr}^{3/2}}
\label{eq:AllRates}
\end{eqnarray}

In the long wavelength limit the most important contributions to decay rates are associated with both longitudinal and transverse phonon scattering into transverse phonons because in this limit $\omega^{3/2}\gg \omega^2$.   This difference is caused by the divergence of  transverse phonons density of states $g_{\rm tr}(\omega)$ at a  frequency  approaching zero 
\begin{eqnarray}
g_{\rm tr}(\omega) \frac{a}{2\pi\hbar}\int dq \delta(\omega-c_{tr}aq^2)= \frac{1}{2\hbar a\sqrt{\Omega_{\rm tr}\omega}}, 
\label{eq:TrDOS}
\end{eqnarray}
in a striking contrast with the constant longitudinal phonon density of states in that limit $g_{\rm l}(\omega) = 1/(\hbar\Omega_{L}a)$. Therefore, there is a bigger chance for phonons  to scatter to a denser transverse band.  The difference between  scattering rates is determined by the difference in densities of states.  

Thus, we evaluated decay rates of both longitudinal and transverse phonons as 
 \begin{eqnarray}
\gamma_{\rm l} =  \gamma_{\rm l,l} +\gamma_{\rm l,tr}= na\left(2\frac{d^2}{\Omega_{\rm l}}\frac{\omega^{2}}{\Omega_{\rm l}^{2}}+\frac{b^2}{\Omega_{\rm l}}\frac{\omega^{3/2}}{\Omega_{\rm tr}^{3/2}}\right),
\nonumber\\
\gamma_{\rm tr}= \gamma_{\rm tr,l}+\gamma_{\rm tr,tr} = na\left(2\frac{b^2}{\Omega_{\rm tr}}\frac{\omega^{2}}{\Omega_{\rm l}^{2}} + \frac{f^2}{\Omega_{\rm tr}}\frac{\omega^{3/2}}{\Omega_{\rm tr}^{3/2}}\right). 
\label{eq:LTrRates}
\end{eqnarray}
These rates are used below in Sec. \ref{sub:ThermC} to evaluate the thermal conductivity. 

\subsection{Calculation of thermal conductivity}
\label{sub:ThermC}

Since phonon decay by means of forward and backwards scattering takes place with identical rates, the scattering is isotropic. This means the phonon decay rates, Eq. (16), determine their mean free paths as $l_{\rm l,tr}=v_{\rm l,tr}/\gamma_{\rm l,tr}$, where $v_{\rm l,tr}=d\omega_{\rm l,tr}/dq$ is the phonon transport velocity (remember  that $v_{\rm l}=c_{\rm l}$ and $v_{\rm tr}=2c_{\rm tr}aq$ within the long wavelength limit, see Eq. (\ref{eq:FullH})).  

The phonon's  mean free path determines its  transmission coefficient in Eq. (\ref{eq:kappaDvira}) as $\mathcal{T}(\omega) = e^{-L/l_{\rm l,tr}}$.   This exponential dependence is the consequence of the inevitable Anderson localization of phonons in one dimension with the localization length determined by their mean free path  \cite{Abrahams1979ScThLoc,Gorkov792D}.   Even if the localization length differs by a factor of order of unity from  the mean free path estimated using the Fermi Golden rule in Sec. \ref{sub:FGRmod}, our results will have a correct analytical parametric dependence.  

In the limit of a small temperature gradient $T_{R}-T_{L} \ll T_{R}$ (cf. Eq. (\ref{eq:kappaDvira})) the thermal conductivity of the specific phonon band $\mu$ is defined as 
\begin{eqnarray}
\kappa_{\mu} = \frac{k_{B}}{2\pi}L\int\frac{\hbar^2\omega^2e^{-\frac{\hbar\omega}{k_{B}T}}}{(k_{B}T)^2\left(1-e^{-\frac{\hbar\omega}{k_{B}T}}\right)} e^{-L\frac{\gamma_{\rm \mu}(\omega)}{v_{\rm \mu}(\omega)}}d\omega.
\label{eq:kappaDviraGenT}
\end{eqnarray}  
The replacement of summation with integration is justified if the phonon inter-level spacing (e. g.  $2\pi c_{l}/L$ for longitudinal phonons) is less than the thermal energy $k_{B}T$.  We assume that this condition is satisfied. Otherwise, thermal conductivity decreases exponentially with the temperature. 

The main contribution to the thermal conductivity in Eq. (\ref{eq:kappaDviraGenT}) is originated from small frequencies, where the exponential term  is of order of unity.   We formally define the crossover frequencies $\omega_{\rm L,\mu}$ for each band $\mu$ setting this exponent to unity,  i. e.  $v_{\rm \mu}(\omega)/\gamma_{\rm \mu}(\omega_{L})=L$.  Using the dominating contribution of decay rate associated with the phonon scattering to the transverse band we estimate these crossover frequencies for both bands to be 
\begin{eqnarray}
\omega_{\rm L,l}=\frac{1}{(nL)^{2/3}}\Omega_{\rm tr}\left(\frac{\Omega_{\rm l}}{b}\right)^{4/3}, ~~ \omega_{\rm L,tr}=\frac{2}{nL}\Omega_{\rm tr}\left(\frac{\Omega_{\rm tr}}{f}\right)^{2}.
\label{eq:omegL}
\end{eqnarray}
Then two different regimes for thermal conductivity exist depending on the relationship between the thermal energy $k_{B}T$ and the energy of the phonon at the crossover $\hbar\omega_{\rm L,\mu}$.  At low temperature 
$k_{B}T < \hbar\omega_{\rm L,\mu}$ all essential phonons with energy less than the thermal energy propagate ballistically.  In this regime we can approximately set the transmission (exponent) in Eq.  (\ref{eq:kappaDviraGenT}) to unity and evaluate the thermal conductivity as  \cite{2021ballist} 
\begin{eqnarray}
\kappa_{\rm bal} =\frac{\pi^2}{3}k_{B}L\frac{k_{B}T}{2\pi\hbar}.
\label{eq:KappaBal}
\end{eqnarray} 
This contribution  is equal to the product of the conductor length $L$ and the quantum of thermal conductance    $\pi^2k_{B}^2T(6\pi\hbar)$ as derived in Refs.  \cite{1998QuantThermCond,1998HeatTranspRev} and confirmed experimentally in Ref.  \cite{2000NatQuantThermCond}. This result  is universal and independent of the specific band.  

In the opposite limit of relatively high temperatures $k_{B}T > \hbar\omega_{\rm L,\mu}$  we evaluate the integrals in Eq. (\ref{eq:kappaDviraGenT}) analytically, setting $dn/dT=k_{B}/(\hbar\omega)$. Then, leaving only dominating contributions to the phonon decay rates we analytically evaluate  thermal conductivities in the high temperature (long length) limit   for the longitudinal phonons as 
\begin{eqnarray}
\kappa_{\rm l} =\frac{\Gamma(5/3)}{2\pi}k_{B}L\omega_{\rm L,l}=\frac{\Gamma(5/3)}{2\pi}\left(\frac{L}{a}\right)^{1/3}\frac{1}{(na)^{2/3}}k_{B}a\Omega_{\rm tr}\left(\frac{\Omega_{\rm l}}{b}\right)^{4/3},  ~ k_{B}T >\frac{1}{(nL)^{2/3}}\hbar \Omega_{\rm tr}\left(\frac{\Omega_{\rm l}}{b}\right)^{4/3},
\label{eq:KappaSupL}
\end{eqnarray}
and for the transverse phonons as 
\begin{eqnarray}
\kappa_{\rm tr} = \frac{k_{B}}{2\pi}L\omega_{\rm L,tr}=  \frac{1}{\pi na}k_{B}a\Omega_{\rm tr}\left(\frac{\Omega_{\rm tr}}{f}\right)^{2},  ~ k_{B}T > \frac{2}{nL}\hbar\Omega_{\rm tr}\left(\frac{\Omega_{\rm tr}}{f}\right)^{2}.
\label{eq:KappaSupT}
\end{eqnarray} 
The thermal conductivity length dependence given by Eq.  (\ref{eq:KappaSupL}) can be attained by means of increasing the thermal conductor length above the acoustic phonon transport length.  The associated  constraint on the length $L$ takes the form 
\begin{eqnarray}
L > \frac{c_{l}}{\gamma_{\rm l,tr}}\approx \frac{1}{n}\left(\frac{\hbar\Omega_{\rm l}}{k_{B}T}\right)^{3/2}\left(\frac{\Omega_{\rm l}}{b}\right)^{2}. 
\label{eq:LDepLDepL}
\end{eqnarray}

Based on the estimate of the thermal conductivity in Eqs.  (\ref{eq:KappaSupL}),   (\ref{eq:KappaSupT}) we conclude that the main contribution to the thermal conductivity at high temperatures or long lengths   is originated from the  longitudinal phonons, while its size dependence $\kappa \propto L^{1/3}$ is caused by longitudinal to transverse phonon scattering.  Interestingly, the contribution of transverse phonons is size independent.  However,  it is determined by the narrow domain of low frequency phonons propagating ballistically, while typically the length independent thermal conductivity is due to the diffusive transport.  

\begin{figure}%
    \centering
    \subfloat[\centering ]{{\includegraphics[width=7.5cm]{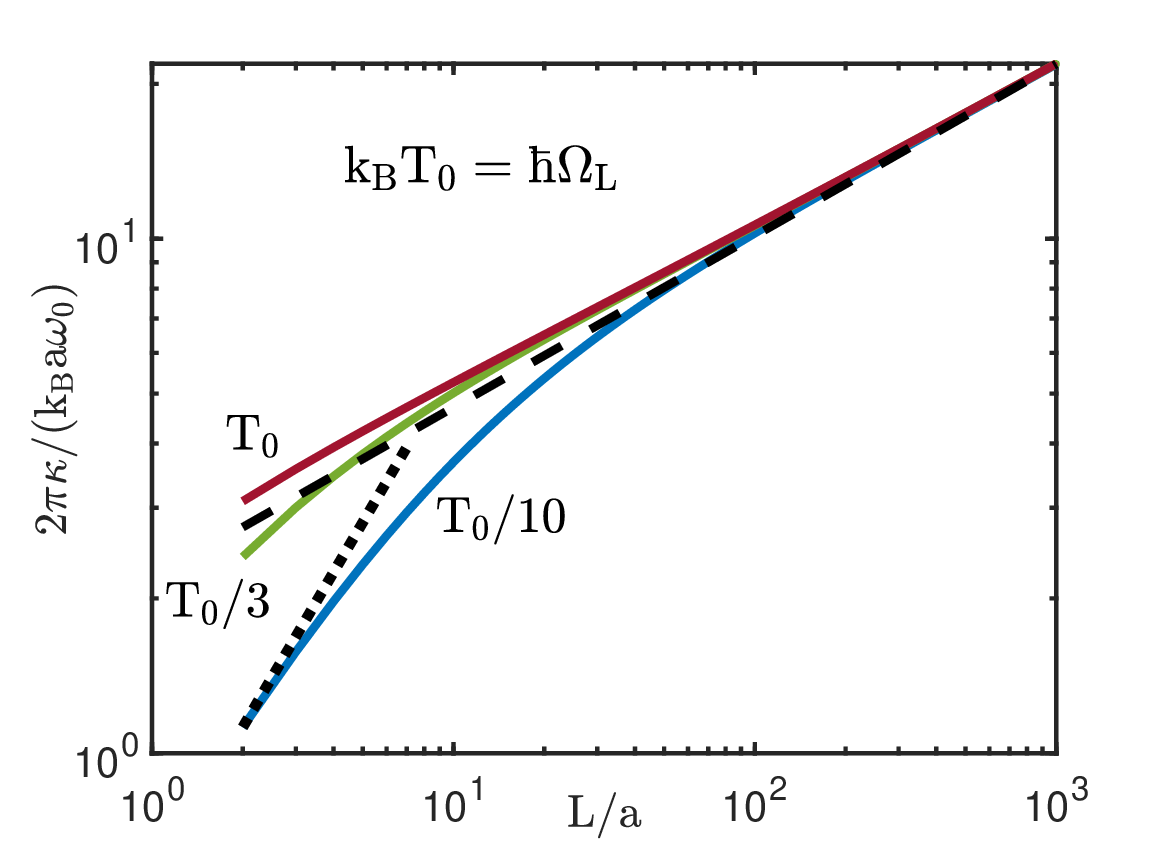} }}%
    \qquad
    \subfloat[\centering ]{{\includegraphics[width=7.5cm]{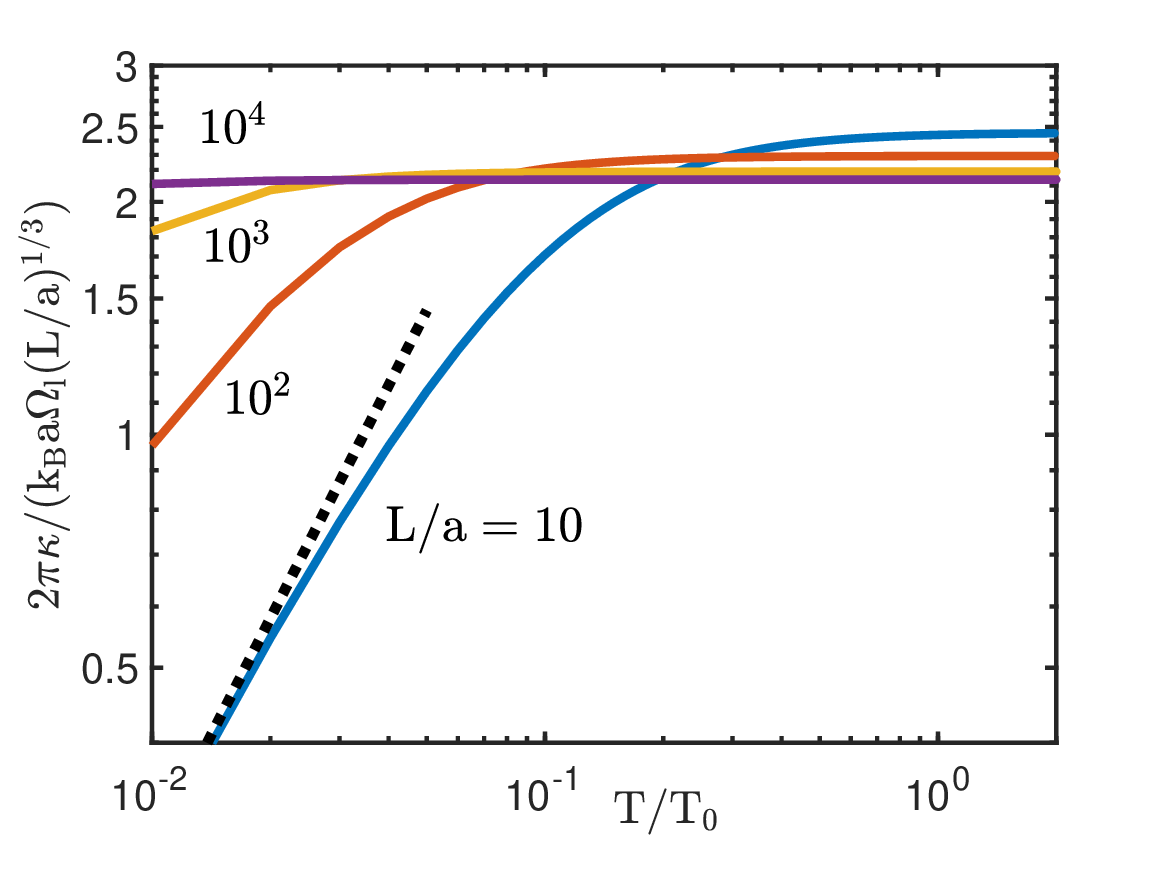} }}%
    \caption{Size  (a)  and temperature  (b) dependencies of thermal conductivity.    Different solid lines in  graph (a) correspond to different temperatures indicated near each graph.  Dotted line shows the linear size dependence corresponding to the ballistic regime,  while dashed line shows $L^{1/3}$ dependence.  In graph (b) the solid lines describe temperature dependence of thermal conductivity for different lengths, while dotted  line shows the linear temperature dependence}%
    \label{fig:Tdepkapp}%
\end{figure}

Using Eq. (\ref{eq:kappaDviraGenT}) we evaluated the  thermal conductivity numerically as a sum of  the longitudinal and transverse thermal conductivities.  The length and temperature dependencies of thermal conductivity are
shown in Fig. \ref{fig:Tdepkapp}.a and b, respectively, for the parameters   chosen as $b=d=f=\Omega_{\rm l}$, $\Omega_{\rm tr}=\Omega_{\rm l}/2$, $n=0.1/a$.  

Both fully ballistic,  Eq.  (\ref{eq:KappaBal}), and temperature independent superdiffusive, Eq.  (\ref{eq:KappaSupL}),  regimes can be clearly identified in both graphs at low temperatures and short lengths or high temperatures and long lengths.   Minor  deviations from the $L^{1/3}$ law are seen  in Fig.  \ref{fig:Tdepkapp}.b  where the thermal conductivities are rescaled by the factor $L^{-1/3}$.  These deviations are caused by  the contribution of transverse phonons.  This contribution  is almost size independent.  It leads to  a minor weakening of $L^{1/3}$ law, that gets less visible at larger sizes.  

In the intermediate temperature domain both size and temperature dependence of thermal conductivity is intermediate  between the low temperature pure ballistic regime $\kappa \propto L\cdot T$ and the high temperature temperature  independent $L^{1/3}$ behavior.   Both somewhat similar and somewhat different behavior has been observed experimentally in different one-dimensional heat conductors as discussed  in Sec.  \ref{sec:Exp}.  

The predicted size dependencies for thermal conductivities of longitudinal and transverse phonons are identical to those found in Ref. \cite{2004TrVibrMD} when using molecular dynamic simulations and mode coupling theory in the one-dimensional defect free model of heat conductor.  In contrast to the present work the phonon scattering there is caused by anharmonic interactions. Therefore, the thermal conductivity rapidly decreases with increasing  temperature.   Thus,  the present work is valid for low temperatures or long lengths (see Eq. (\ref{eq:KappaSupL})), while at very high temperatures the phonon scattering due to anharmonic interactions should dominate. 
The similarity of size dependencies found in  Ref.  \cite{2004TrVibrMD} and here can be explained by assuming that for an anharmonic interaction the fast long wavelength phonons are scattered by the slow, high frequency phonons almost like being scattered by static defects.  At relatively high temperatures such defects naturally emerge  in one-dimensional nonlinear lattices in  the form of metastable discrete breathers \cite{2003DiscreteBreathFlach}. 

Below in Sec.  \ref{sec:Fence} we confirm the relevance of theory prediction using the simple  chain model with defects shown in Fig. \ref{fig:2dModel}.

\subsection{Torsional phonon contribution} 
\label{sub:Tors}

Torsional phonon spectrum is similar to that  of longitudinal phonons.  They can also be scattered to transverse phonons by defects.  The associated interaction  takes the form  $G_{\rm tors}(\partial \theta/\partial x)\times (\partial^2 u_{y}/\partial x^2)$, where $\theta$ is a torsional angle displacement.  This interaction requires violation of inversion and reflection symmetries, which should take place in realistic polymer molecules due to structural defects like kinks.   Consequently,  we expect that the proposed torsional to transverse phonon scattering leads to the  identical  phonon decay rate frequency dependence given by Eq. (\ref{eq:LTRDec})  and,  consequently,  to the  identical thermal conductivity size dependence $\kappa \propto L^{1/3}$ for torsional phonons. 

\subsection{Two dimensions} 
\label{sub:2D}

It is straightforward to extend the above consideration to the thermal conductivity of two-dimensional materials where the transverse phonons with displacements perpendicular to the material plane also possess a quadratic spectrum.   The longitudinal to transverse phonon scattering rate, evaluated similarly to Eq. (\ref{eq:LTRDec}), is characterized by the frequency dependence $\gamma_{2D}(\omega) \propto \omega^2$. Similarly to Ref. \cite{2004TrVibrMD} this dependence can result in logarithmic size dependence of thermal conductivity due to phonons propagating in a diffusive manner but with the diffusion coefficient growing by increasing the wavelength.  Since an accurate analysis of thermal conductivity needs an accurate solution of the complicated diffusion problem involving phonon scattering between longitudinal and transverse bands (cf.  Ref.  \cite{ab97PhTrPhysRep} where the effect of phonon ``absorption'' and reemission by quasi-local modes was addressed), we postpone it for  future work.   

Three-dimensional materials do not have the transverse acoustic phonon band with quadratic spectrum. 

\section{Phonon scattering within the ``fence'' model}
\label{sec:Fence} 

\subsection{Fence model}
\label{sub:FenceMod}

\begin{figure}
\includegraphics[scale=0.5]{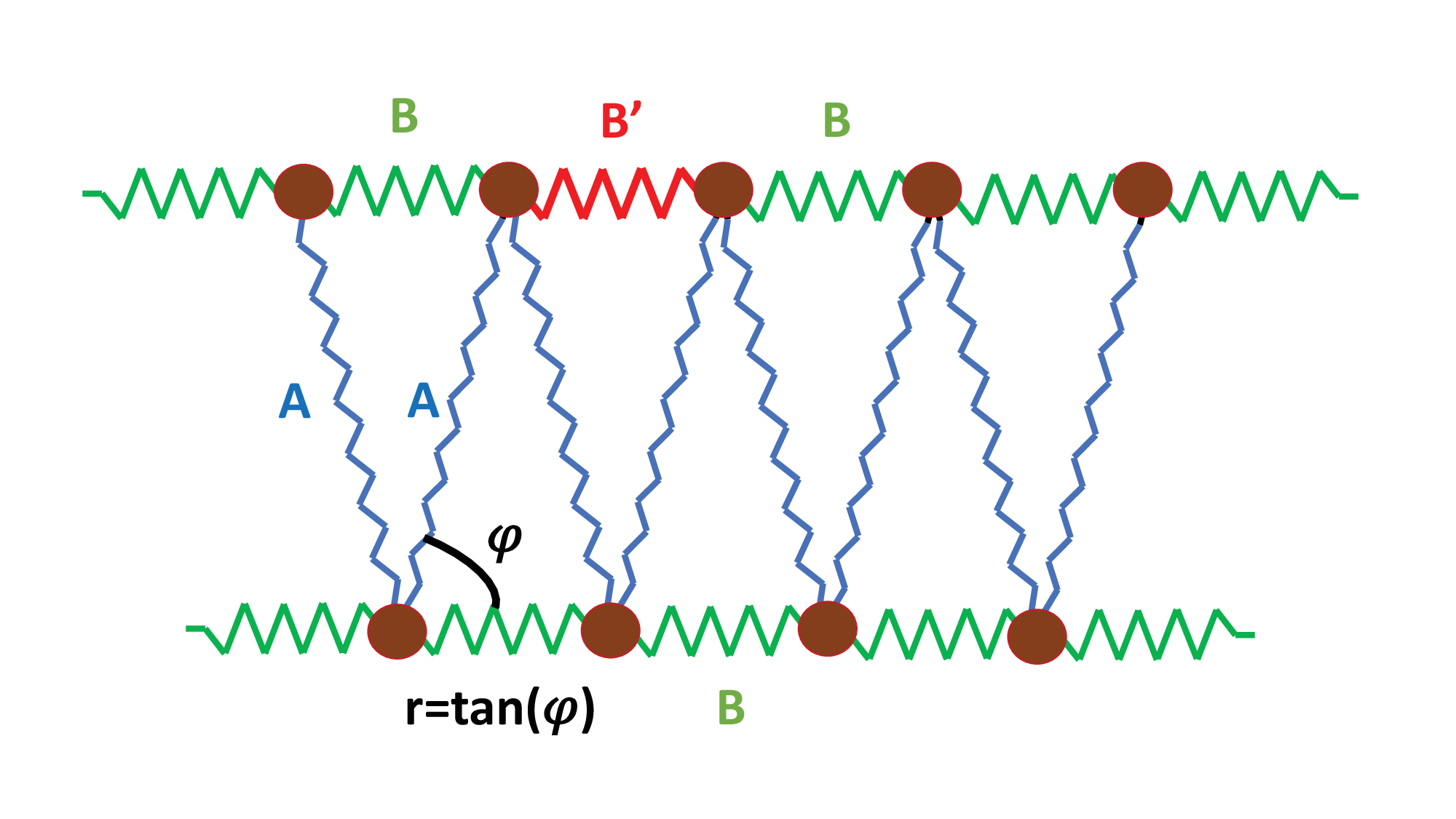} 
\caption{The fence model of polymer  chain  containing defect coupling longitudinal and transverse phonons.  Springs connect neighboring and nearest neighboring atoms shown the brown circles.  Nearest neighbor interactions are all identical and characterized by the force constant $A$, while next-neighbor couplings are characterized by the force constant $B$ except for the two defects sites in the top coupled with the strength $B'$ (shown by the red color online).}
\label{fig:2dModel}
\end{figure}

To illustrate the effect of longitudinal to transverse phonon scattering and prove its existence,  we use the toy ``fence" model of chain with the springs connecting nearest and next to nearest neighbors as shown in Fig.  \ref{fig:2dModel}.  In our consideration we ignore the third dimension which is not needed for the process of interest.  The minimum energy chain geometry shown in Fig.  \ref{fig:2dModel} emerges naturally in two dimensions in any model including the only nearest and next neighbor interactions $U_{1}(R)$ and $U_{2}(R)$ depending only on interatomic distances if these interaction possess energy minima at interatomic distances $R_{1}$ and $R_{2}$, respectively, such  that $2R_{1} > R_{2}$. In this case the nearest neighboring distances are equal to $R_{1}$ and the next neighboring distances are equal to $R_{2}$.  In this geometry,  the angle $\varphi$ between the chain axis $x$ and the vector connecting the nearest neighbors is defined as $\sin(\varphi) = R_{2}/(2R_{1})$.   The model of chain vibrations including both the nearest and next neighbor interactions was suggested earlier in Ref.  \cite{1970LEE1DModelofVibrWithnnextint} under assumption $R_{2}=2R_{1}$ corresponding to the linear chain geometry.  The  similar model involving transverse displacements was used in Ref.  \cite{2004TrVibrMD} but with the  next neighbor interaction replaced with the angular dependent potential energy (see also Refs. \cite{19762DversionLEE,2010TransvVibr1D}), which is the natural consequence of angular dependent bond interactions.    The proposed model is more complicated because  it possesses two-dimensional  geometry compared to the earlier models focused on  purely one-dimensional chains.  However, it seems to not be  too  far from  realistic  molecular chains like polyethylene. Therefore, this model can serve as a paradigmatic model  capable of treating  longitudinal and transverse phonons  in real molecules.  To the best of our knowledge, this relatively simple model was not previously considered.  We did not find any similar consideration in spite of an extensive check of existing literature.

The Hamiltonian of the model represented by Fig. \ref{fig:2dModel} in the harmonic approximation takes the form 
\begin{eqnarray}
\widehat{H} = \frac{1}{2}\sum_{n}\left(\mathbf{p}_{n}^2+A(x_{n+1}-x_{n}-r(-1)^n(y_{n+1}-y_{n}))^2+B_{n}(x_{n+2}-x_{n})^2\right), 
\label{eq:HFence}
\end{eqnarray}
where the mass of atoms is set to unity, $x_{n}$, $y_{n}$ are longitudinal and transverse displacements of $n^{th}$ atom, $r=\tan(\varphi)$ and $\varphi$ is the angle between vectors connecting nearest and next neighbors (see Fig.  \ref{fig:2dModel}) and  $A$ (actually $A\sqrt{1+r^2}$) and $B_{n}$ are force constants responsible for the nearest and next neighboring interactions., respectively.  The force constant $B_{n}$ is equal to $B$ everywhere but at rare defect sites, where it is equal to $B'$.   We use notations $A$ and $B$ for the force constants, while we used the same letters for  the other  parameter notations in the previous section.  Remember  that they have different meanings. There is a finite number of letters,  so some repetitions are unfortunately  unavoidable. 

The specific form of the potential energy  in Eq. (\ref{eq:HFence}) is caused by its dependence on the absolute value of interatomic distance.  In this case the second order expansion of the potential energy $\sum_{\mu,\nu}\partial^2U(r_{ij})/(\partial x_{\mu}\partial x_{\nu})(u_{i\mu}-u_{j\mu})(u_{i\nu}-u_{j\nu})/2$ ($\mu$,  $\nu$ are coordinate indices $x$ and $y$,   $u_{i\mu}$ is the displacement of a site $i$ in a direction $\mu$) takes the form 
$$
\frac{1}{2}\frac{d^2U}{dr^2}\left((\mathbf{u}_{i}-\mathbf{u}_{j})\mathbf{n}_{ij}\right)^2, ~ \mathbf{n}_{ij}=\frac{\mathbf{r}_{ij}}{r_{ij}},
$$
used in Eq.  (\ref{eq:HFence}).  Deriving this expression we used the condition $dU/dr=0$ for all distance dependent potential energies, since in our model (see Fig. \ref{fig:2dModel}) both neighboring and next neighboring interatomic distances realize the minimum of the corresponding potential energy.  A defect is introduced for the next-neighbor interaction since in that position it violates both reflection and inversion symmetries (see Fig.  \ref{fig:2dModel}).  

Below, in Sec.  \ref{sub:FenceMod}, we define the normal modes of the Hamiltonian Eq. (\ref{eq:HFence}) in the absence of defects and describe their scattering by defects in terms of the transmission coefficient in Sec.  \ref{sub:TrRefl}.  It turns out that transmission and reflection frequency dependencies are consistent with the predictions of the phenomenological model developed in Sec.  \ref{sub:LonWvModel}.

\subsection{Normal Modes}
\label{sub:FenceMod}

The Hamiltonian Eq.  (\ref{eq:HFence}) is periodic with the period $2a$ equal to the distance between next neighbors as shown in Fig.  \ref{fig:2dModel}.  This period can be reduced twice by the coordinate transformation $y_{n}\rightarrow (-1)^ny_{n}$ that modifies the Hamiltonian as 
\begin{eqnarray}
\widehat{H} = \frac{1}{2}\sum_{n}\left(\mathbf{p}_{n}^2+A(x_{n+1}-x_{n}+r(y_{n+1}+y_{n}))^2+B_{n}(x_{n+2}-x_{n})^2\right). 
\label{eq:HFenceMod}
\end{eqnarray}
In the  absence of defects  ($B_{n}=B$) we can use the Fourier transformed coordinates and momenta, assuming for the sake of simplicity periodic boundary conditions. Then Eq.  (\ref{eq:HFenceMod}) takes the form 
\begin{eqnarray}
\widehat{H} = \sum_{q>0}\left(\mathbf{p}_{q}\mathbf{p}_{-q}+x_{q}x_{-q}\left(2A(1-\cos(qa))+2B(1-\cos(2qa))\right)\right.
\nonumber\\
\left.+ i2Ar\sin(qa)(x_{q}y_{-q}-x_{-q}y_{q})+2Ar^2(1+\cos(qa))y_{q}y_{-q})\right). 
\label{eq:HFenceModq}
\end{eqnarray}

\begin{figure}
\includegraphics[scale=0.5]{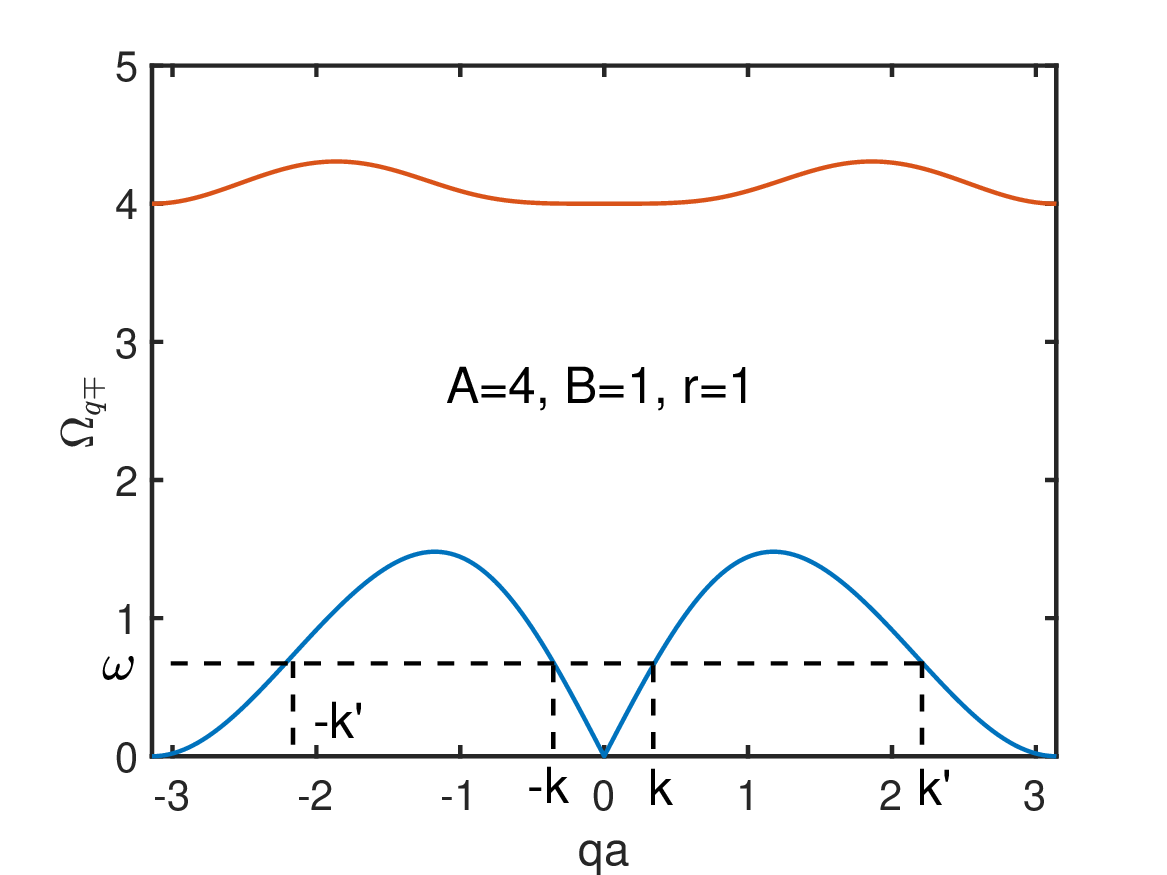} 
\caption{The spectrum of normal modes of the ``Fence'' Hamiltonian, Eq.  (\ref{eq:HFence}, as given by Eq. (\ref{eq:HfenceNormModes}). The model parameters are shown in the graph.  The bottom (blue in colored version) graph describes acoustic bands, while the top graph describes the optical bands.   The frequency of acoustic band approaches $0$ at zero wavevectors as $\omega_{q} \propto |q|$ for longitudinal modes and as  $\omega_{q} \propto (\pi-|q|)^2$ for $q\rightarrow \pm \pi$ for transverse modes. }
\label{fig:FenceSpectra}
\end{figure}

The further transition to the normal coordinates is performed using the transformation 
\begin{eqnarray}
x_{q}=\cos(\phi_{q}/2)u_{q}+i\sin(\phi_{q}/2) v_{q}, ~ y_{q}=-i\sin(\phi_{q}/2)u_{q}+\cos(\phi_{q}/2) v_{q}, 
\nonumber\\
 \tan(\phi_{q})=\frac{2Ar\sin(qa)}{A(1-\cos(qa))-Ar^2(1+\cos(qa))+B(1-\cos(2qa)}. 
\label{eq:uvtransf}
\end{eqnarray}
This transformation leads to independent   normal modes of acoustic and optical phonons characterized by the  coordinates $u_{q}$ and $v_{q}$, respectively.  The Hamiltonian then takes the form   
 \begin{eqnarray}
 \widehat{H}=\sum_{q>0} \left(p_{qu}p_{-qu} +\Omega_{q-}^2u_{q}u_{-q}\right)+ \left(p_{qv}p_{-qv} +\Omega_{q+}^2v_{q}v_{-q}\right),
 \nonumber\\
 \Omega_{q\mp}^2=A(1-\cos(qa))+Ar^2(1+\cos(qa))+B(1-\cos(2qa))
\nonumber\\ 
 \mp \sqrt{\left(A(1-\cos(qa))+Ar^2(1+\cos(qa))+B(1-\cos(2qa))\right)^2+4A^2r^2\sin(qa)^2}. 
\label{eq:HfenceNormModes}
\end{eqnarray}
The spectra of the normal modes versus the wavevector are shown in Fig. 2 for the parameters chosen as $A=4$,  $B=1$, $r=1$.  The negative  sign in the frequency definition in Eq. (\ref{eq:HfenceNormModes}) corresponds to the two acoustic bands, while the positive sign corresponds to the two optical bands of the original model in Eq.  (\ref{eq:HFence}).  Wavevectors from the domain $(-\pi/(2a), \pi/(2a))$ approximately correspond to longitudinal modes, while the remaining domain $(-\pi/a, -\pi/(2a))$ and $(\pi/(2a), \pi/a)$ is related to the transverse like modes.  The  wavevectors for the two atomic elementary cell should be redefined for transverse modes as $q \rightarrow q+\pi$ for $q<0$ and $q \rightarrow q-\pi$ for $q<0$.  

The frequency $\Omega_{q-}$  in Eq. (\ref{eq:HfenceNormModes}) approaches $0$ for $q\rightarrow 0$ as $\Omega_{q-}=|q|a\sqrt{2B}$ as in the acoustic band with only next neighbor interactions.  No stretching of the nearest neighbor bonds takes place for acoustic vibrations since the vertical displacements of atoms keeps the  nearest neighbor distances almost unchangeable.  In the limit $q'\rightarrow 0$ ($q'=\pi/a-|q|$) the frequency  $\Omega_{q-}$ approaches $0$ as $\Omega_{q-}=q^2a^2\sqrt{B/2}r$ in full accord with the expected behavior of  transverse acoustic modes.  Below in Sec. \ref{sub:TrRefl} we examine scattering of these modes by a single defect, formed due to the modified force constant $B$.  


\subsection{Transmission and reflection  in the presence of a single defect}
\label{sub:TrRefl}


Here, we study the transmission and reflection of phonons by a single defect located at site $n=0$, where we set the next neighbor force constant in the Hamiltonian Eq. (\ref{eq:HFenceMod}) to be equal $B'\neq B$.  The transmission coefficient $T_{\rm l,l}$ tells us about the probability  of the longitudinal wave packet to successfully pass a single defect.   The probability to pass $k$ defects without reflection takes the form $T_{\rm l}^{k}$ for non-correlated positions of defects.  Consequently, the ballistic transport  intensity drops with the distance $x$ approximately as $e^{-nx\ln(T_{\rm l,l})}$ which leads to the estimate of the mean free path $l$ and the phonon decay rate $\gamma$ as (remember that $v$ is the phonon group velocity) 
\begin{eqnarray}
l=\frac{1}{n\ln(1/T_{\rm l,l})}, ~ \gamma = \frac{v}{l}=nv\ln(1/T_{\rm l,l})\approx nv(1-T_{\rm l,l})). 
\label{eq:meanfrpth}
\end{eqnarray}
The latter approximate identity is valid in a long-wavelength limit under consideration, where $1-T_{\rm l,l} \ll 1$.

\begin{figure}%
    \centering
    \subfloat[\centering ]{{\includegraphics[width=7.5cm]{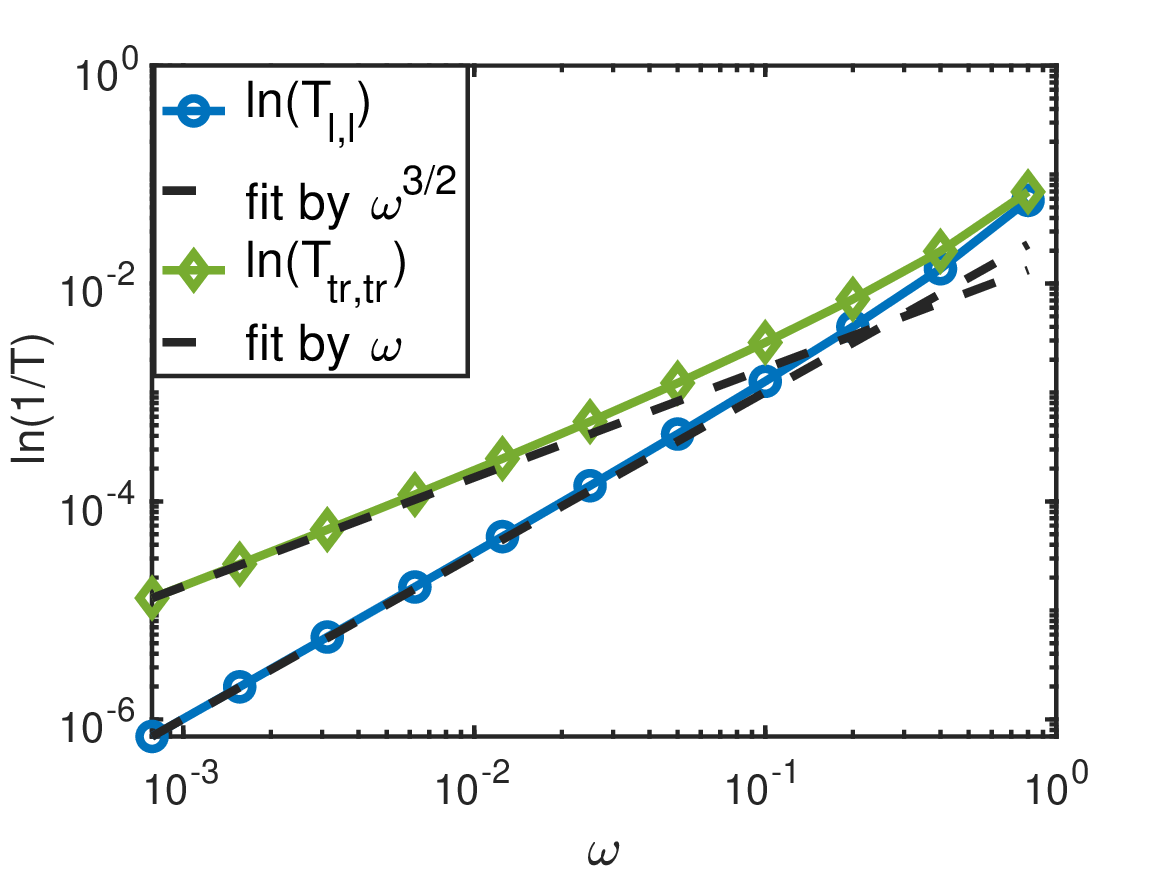} }}%
    \qquad
    \subfloat[\centering ]{{\includegraphics[width=7.5cm]{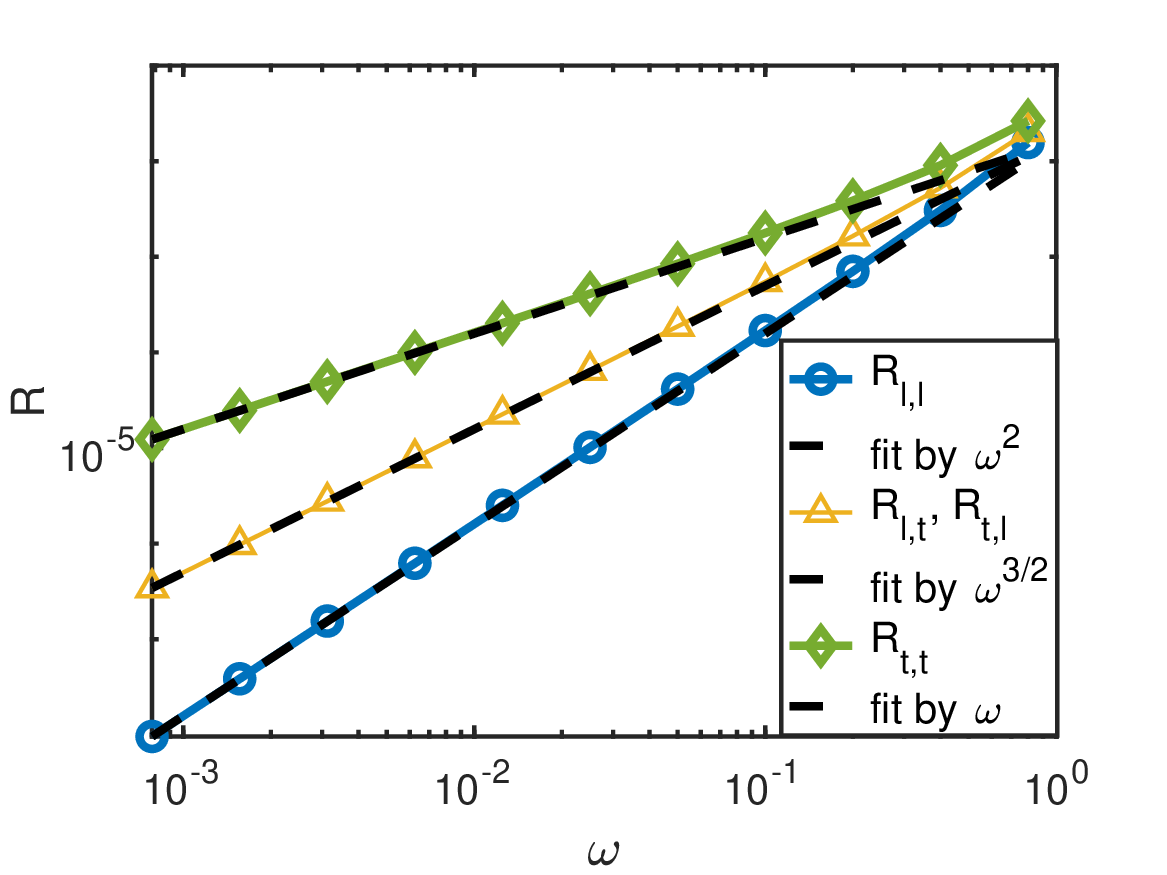} }}%
    \caption{Frequency dependence of  (a) transmission logarithm  (b) reflections for different incident and outgoing phonons.  Dashed lines indicate the analytical predictions for the long-wavelength limit obtained in Sec.  \ref{sec:LongWvlPhen}.  The results for transmissions $T_{r,l}$ and $T_{l,r}$ are not shown, since they are almost identical to corresponding reflections. }%
    \label{fig:TrRefl}%
\end{figure}

We assume that the incident longitudinal phonon with the frequency $\omega$ and the wavevector $k$ has a unit amplitude.   Then, its scattering by defect results in the formation of two passed waves with wavevectors $k$ and $-k'$, and two reflected waves with the wavevectors $-k$ and $k'$,  where the transverse phonons with the wavevector $\pm k'$ possess the same frequency $\omega$  as shown in  Fig. \ref{fig:FenceSpectra}.   The coordinates $x_{n}$  and $y_{n}$ well before and after the defect ($n<0$,  $|n| \gg 1$ or  ($n>0$,  $|n| \gg 1$) should behave as 
\begin{eqnarray}
x_{n} = \begin{cases}
u_{k}(e^{ikn}+r_{\rm l}e^{-ikn}) +r_{\rm tr}u(k')e^{ik'n}\sqrt{\frac{v_{l}}{v_{\rm tr}}}, ~ n\ll -1,  \\
t_{\rm l}u(k)e^{ikn}+t_{\rm tr}u(k')e^{-ik'n}\sqrt{\frac{v_{\rm l}}{v_{\rm tr}}}, ~ n \gg 1, 
\end{cases}
\nonumber\\ 
y_{n} = \begin{cases}
v_{k}(e^{ikn}+r_{\rm l}e^{-ikn}) +r_{\rm tr}v(k')e^{ik'n}\sqrt{\frac{v_{l}}{v_{\rm tr}}}, ~ n\ll -1,  \\
t_{\rm l}v(k)e^{ikn}+t_{\rm tr}v(k')e^{-ik'n}\sqrt{\frac{v_{\rm l}}{v_{\rm tr}}}, ~ n \gg 1, 
\end{cases}
\nonumber\\ 
v_{\rm l} =\left|\frac{d\Omega_{k-}}{dk}\right|, ~ v_{\rm tr} =\left|\frac{d\Omega_{k'-}}{dk'}\right|. 
\label{eq:waveasympt}
\end{eqnarray}
The factor containing the square root of the ratio of the two group velocities accounts for the conservation of energy flux during scattering.  Transmissions and reflections are defined using squared absolute values of the wave amplitudes as 
\begin{eqnarray}
T_{\rm l,l}=|t_{\rm l}|^2, ~ T_{\rm l,tr}=|t_{\rm tr}|^2,  ~ R_{\rm l,l}=|r_{\rm l}|^2, ~ R_{\rm l,tr}=|r_{\rm tr}|^2,
\label{eq:TransmRefl}
\end{eqnarray}
where $T_{\rm \mu\nu}$ stands for transmissions and   $R_{\rm\mu\nu}$ stands for reflections. The first subscript denotes the incident wave and the second subscript denotes the transmitted or reflected wave, which are longitudinal for the subscript ``l'' and transverse for the subscript ``tr''.  The transmissions and reflections for the transverse wave scattering are defined identically.   This definition should satisfy the energy conservation law in the form 
\begin{eqnarray}
T_{\rm l,l}+T_{\rm l,tr}+R_{\rm l,l}+R_{\rm l,tr}=1. 
\label{eq:EnCons}
\end{eqnarray}

In addition to the general connection between the incident phonon transmission and its decay rate Eq. (\ref{eq:meanfrpth}), there is also the connection between the transmission and reflection coefficients and phonon decay rates in corresponding channels.   Transmission ($T_{\rm \mu,\nu}$ with  $\mu\neq \nu$) and reflection ($R_{\rm \mu,\nu}$) coefficients, represent the probabilities $P_{\rm \mu,\nu,\pm}$ of the incident phonon forward scattering ($P_{\rm \mu,\nu,+}=T_{\rm \mu,\nu}$) or backwards scattering  ($P_{\rm \mu,\nu,-}=R_{\rm \mu,\nu}$),  which are connected to the corresponding phonon decay rates as  $P_{\rm \mu,\nu,\pm} \approx \gamma_{\rm \mu,\nu,\pm}/(nv_{\rm\mu})$.  Using  this relationship and the frequency dependent phonon decay rates estimated in Sec. \ref{sec:LongWvlPhen}, we predict the transmission and reflection coefficient frequency dependencies 
\begin{eqnarray}
R_{\rm l,l} \propto \omega^2, ~ R_{\rm l,tr} \approx R_{\rm tr,l} \approx T_{\rm l,tr}\approx T_{\rm tr,l}\propto \omega^{3/2}, ~ R_{\rm tr,tr} \propto \omega. 
\label{eq:TrReflFreq}
\end{eqnarray}

Below, we evaluate the transmission and reflection  coefficients numerically and compare the associated phonon decay rate with the predictions using the  Fermi Golden rule estimates reported in Sec.  \ref{sec:LongWvlPhen}.  We determined the transmission and reflection coefficients using the generalized transfer matrix method  for
tight-binding models developed in Ref.  \cite{2016GeneralizedTransfMatr1D}.   It is important to note that in accordance with Ref.   \cite{2016GeneralizedTransfMatr1D}, the solutions decreasing exponentially with the distance from the defect should be added to Eq.  (\ref{eq:waveasympt})  on both sides of the defect, in addition to the running waves.  Since they decrease at much shorter lengths  compared to the estimated mean free path they do not affect the vibrational energy transport at long wavelengths.  

The results for logarithms of transmission coefficients for incident longitudinal and transverse waves vs.  the incident phonon frequency are shown in Fig.  \ref{fig:TrRefl}.a for the scattering by a single defect characterized by the next neighbor force constant $B'=2B$.   In the low frequency limit where $\ln(T_{\rm\mu\mu})\approx 1-T_{\rm\mu\mu}=R_{\rm\mu\mu}+T_{\rm\mu\mu'}+R_{\rm\mu\mu'}$ ($\mu'\neq \mu$) their frequency dependencies  are consistent with the frequency dependencies of  dominating  scattering channels $R_{\rm l,tr}\propto \omega^{3/2}$ for longitudinal phonons and $R_{\rm tr,tr}\propto  \omega$ for transverse phonons.  Frequency dependencies of reflection or transmission coefficients in the specific scattering channels are shown in  Fig.  \ref{fig:TrRefl}.b.  These dependencies are perfectly consistent with the predictions of Sec.  \ref{sec:LongWvlPhen} expressed by Eq.  (\ref{eq:TransmRefl}).  Moreover they better follow the predicted power law dependencies compared to the total reflections  shown in Fig.  \ref{fig:TrRefl}.a since several contributions characterized by different power law dependencies are added there.  

We also examined the defect formed due to  a modified  constant $A$ and found almost non-detectable  scattering  within the long-wavelength limit.  Since the force constant $A$ does not enter any acoustic phonon spectra within the long-wavelength limit, this is not very surprising.  Also, the modification  in the ``fence" slope $r$ led to weaker scattering of longitudinal phonons compared to that induced by modification of  the force constant $B$, because   it does not enter the longitudinal phonon spectrum as well.   Therefore, we discard  these effects as minor compared to that of the modification of the constant $B$.  We also verified that the mass defect  leads to the  Rayleigh scattering in accordance  with the discussion in Sec. \ref{sub:LonWvModel}.

\section{Discussion of the Experiments}
\label{sec:Exp}

Here, we briefly discuss the existing measurements of length dependence of thermal conductivity in one-dimensional conductors and the connection of the present theory to these experiments.  

 To the best of our knowledge, there is a single experiment reporting thermal conductivity in the molecular system  consisting of  poly (methyl acrylate) (PMA) chains  connecting spherical SiO$_2$ nanoparticles  \cite{2024prlnanopartonpolymsuperdiff}. The  $L^{1/2}$ length dependence of thermal conductivity was  discovered there.  This dependence is the natural outcome of the one-dimensional  Rayleigh scattering of longitudinal phonons by defects within the longitudinal acoustic band \cite{LEBOWITZ00,20011DGlass}
 as was  discussed in the introduction. This  dependence  disagrees with our predictions.  To address this discrepancy, we can only assume that the longitudinal to transverse phonon scattering is absent or very weak in this specific system.  Covalent bonds responsible for the structure of polymer are very strong,  so there can be practically no structural defects like kinks modifying those bonds in the polymer molecules under consideration.     The Rayleigh scattering still exists there because of  isotope defects which always exist in any polyatomic molecule. 

In many experiments   involving  nanowires the dependence $L^{1/3}$ has been observed.    It is remarkable that this dependence was discovered in Si nanowires at the very low temperature of $4$K  \cite{2018SiNanowLdepThCond,2017BallTr4KSiNanow033},  which  is definitely  low enough (cf.  Eq. (\ref{eq:AreaConstr})) to consider the wire as truly one-dimensional.  However, the measurements of the temperature dependence of thermal conductivity within a close temperature domain (e. g. between $4$K and $10$K) are highly desirable  to confirm or disprove a lack of temperature dependence  suggested in the present work.

 Other experiments involving nanowires   \cite{2019IncrThCondSiGe,2024IncrThCondSiCarbide,2021SuperDiffN033}  performed at higher  temperatures  also report a thermal conductivity size dependence close to $L^{1/3}$. Yet, in these experiments thermal conductivity increases with the temperature for the nanowires with small diameter, where this length dependence is observed.  This is not consistent with the present theory in the long length or high temperature limits and is  even less consistent with other theories considering anharmonic interactions, since according to these theories thermal conductivity should decrease with the temperature.   It is possible that the observations are made within the intermediate temperature domain where some temperature dependence remains as shown in Fig. \ref{fig:Tdepkapp}.  Also the generalization of the present theory involving all phonon bands  can possibly explain the observed behavior.  This generalization is beyond the scope of the present work. 

It is presumably quite natural to expect the proposed length dependence  $L^{1/3}$ to be observed in the suspended nanostructures made of silicon nitride  similar to those used to measure thermal conductivity quantum as reported in Ref.  \cite{2000NatQuantThermCond}.  This regime can be attained by using similar nanostructures of identical tranverse sizes and longer length at the same temperature. If the sample length will be made longer than the thermal phonon transport length, Eq. (\ref{eq:LDepLDepL}), the thermal conductivity should become temperature independent and increase with the thermal conductor length as $L^{1/3}$ in accord with Eq. (\ref{eq:KappaSupL}).

\section{Conclusion}
\label{sec:Concl}

We investigated thermal conductivity of one-dimensional materials  in the presence of  defects promoting  the scattering of longitudinal phonons to transverse phonons and backwards.  We show  that this scattering leads to phonon decay with the rate $\gamma$ scaling with the frequency as $\omega^{3/2}$.   This scattering  leads to  the thermal conductivity increasing with the conductor size $L$.   Namely, the thermal conductivity is determined by the tiny fraction ($L^{-2/3}$) of low frequency phonons propagating ballistically. It shows distance dependence $\kappa \propto L^{1/3}$.   

This scattering is stronger compared to the Rayleigh scattering withing the longitudinal phonon band. The latter scattering results in the phonon decay rate $\gamma\propto \omega^{2}$ (Rayleigh scattering in one dimension) and thermal conductivity $\kappa \propto L^{1/2}$ \cite{LEBOWITZ00,20011DGlass,Lebowitz15}, which should be changed  to $L^{1/3}$ dependence due to dominating scattering of longitudinal phonons to transverse phonons.   

The generalization of theory to two-dimensional heat conductors  leads to the logarithmic dependence of thermal conductivity  on the size of the conductor.  The full description of thermal energy transport in two dimensions affected by the longitudinal to transverse phonon scattering requires accurate analysis of phonon diffusion, which is beyond the scope of the present paper.  There is no transverse mode with quadratic spectrum in three dimensions. 

The longitudinal to transverse phonon scattering is significant only for structural defects while for mass defects (e. g. due to isotopes) the Rayleigh scattering should be dominant. 

 The size dependence of thermal conductivity $\kappa \propto L^{1/3}$ was derived earlier as the consequence of hydrodynamic interactions in one dimensional momentum conserving systems  \cite{Turbulence02,2015SuperHeatDiffMomCons} and confirmed by molecular dynamics simulations \cite{2004TrVibrMD}.  This behavior is the outcome of   classical anharmonic interactions.  Consequently,  the associated thermal conductivity should decrease with the temperature because anharmonic scattering gets stronger with increasing the number of phonons.   This is in contrast with our proposed mechanism lacking any temperature dependence at sufficiently high temperature.

The measurements of thermal conductivity size dependence in one-dimensional materials often result in  its $L^{1/3}$ length dependence.  However,  in most of experiments thermal conductivity increases with the temperature  which is not consistent with the present theory,  which predicts no temperature dependence and is even less consistent with the earlier work based on anharmonic interactions that predict the reduction of thermal conductivity with increasing temperature.   We hope that the generalization of the present theory involving all phonon bands  can explain the observed temperature and size dependencies simultaneously. 

Very strong coupling of transverse and longitudinal phonons  should emerge in polymer molecules   due to kinks affecting the direction of the chain axis.   Therefore, in such molecules the longitudinal to transverse phonon scattering should be very significant.  The molecular dynamics simulations of the thermal conductivity of molecules with kinks indeed show substantial reduction of thermal energy transport due to such defects    \cite{ThermCondKinks2019Xuhui}.  The Fourier law found there is not consistent with our consideration.   Anharmonic interactions missed in our consideration can also be significant for such defects.  

Our predictions are relevant at low temperatures, while at higher temperatures anharmonic interaction should dominate.  Based on the present findings we expect that anharmonic interaction involving transverse modes is  significant in one-dimensional conductors at high temperatures as well.

\begin{acknowledgements}

This work was supported by the National Science Foundation
(Grant No. CHE-2201027). The author acknowledge Igor Rubtsov for stimulating discussions, and Saanvi Nair,   Michael Flynn and Petra Radmanovic for many useful critical comments.   

\end{acknowledgements}


\bibliography{Vibr1}

\begin{thebibliography}{69}%
\makeatletter
\providecommand \@ifxundefined [1]{%
 \@ifx{#1\undefined}
}%
\providecommand \@ifnum [1]{%
 \ifnum #1\expandafter \@firstoftwo
 \else \expandafter \@secondoftwo
 \fi
}%
\providecommand \@ifx [1]{%
 \ifx #1\expandafter \@firstoftwo
 \else \expandafter \@secondoftwo
 \fi
}%
\providecommand \natexlab [1]{#1}%
\providecommand \enquote  [1]{``#1''}%
\providecommand \bibnamefont  [1]{#1}%
\providecommand \bibfnamefont [1]{#1}%
\providecommand \citenamefont [1]{#1}%
\providecommand \href@noop [0]{\@secondoftwo}%
\providecommand \href [0]{\begingroup \@sanitize@url \@href}%
\providecommand \@href[1]{\@@startlink{#1}\@@href}%
\providecommand \@@href[1]{\endgroup#1\@@endlink}%
\providecommand \@sanitize@url [0]{\catcode `\\12\catcode `\$12\catcode
  `\&12\catcode `\#12\catcode `\^12\catcode `\_12\catcode `\%12\relax}%
\providecommand \@@startlink[1]{}%
\providecommand \@@endlink[0]{}%
\providecommand \url  [0]{\begingroup\@sanitize@url \@url }%
\providecommand \@url [1]{\endgroup\@href {#1}{\urlprefix }}%
\providecommand \urlprefix  [0]{URL }%
\providecommand \Eprint [0]{\href }%
\providecommand \doibase [0]{http://dx.doi.org/}%
\providecommand \selectlanguage [0]{\@gobble}%
\providecommand \bibinfo  [0]{\@secondoftwo}%
\providecommand \bibfield  [0]{\@secondoftwo}%
\providecommand \translation [1]{[#1]}%
\providecommand \BibitemOpen [0]{}%
\providecommand \bibitemStop [0]{}%
\providecommand \bibitemNoStop [0]{.\EOS\space}%
\providecommand \EOS [0]{\spacefactor3000\relax}%
\providecommand \BibitemShut  [1]{\csname bibitem#1\endcsname}%
\let\auto@bib@innerbib\@empty
\bibitem [{\citenamefont {Choy}(1977)}]{1977ChoyReviewPolymerThCond}%
  \BibitemOpen
  \bibfield  {author} {\bibinfo {author} {\bibfnamefont {C.L.}\ \bibnamefont
  {Choy}},\ }\bibfield  {title} {\enquote {\bibinfo {title} {Thermal
  conductivity of polymers},}\ }\href {\doibase
  https://doi.org/10.1016/0032-3861(77)90002-7} {\bibfield  {journal} {\bibinfo
   {journal} {Polymer}\ }\textbf {\bibinfo {volume} {18}},\ \bibinfo {pages}
  {984--1004} (\bibinfo {year} {1977})}\BibitemShut {NoStop}%
\bibitem [{\citenamefont {Shen}\ \emph {et~al.}(2010)\citenamefont {Shen},
  \citenamefont {Henry}, \citenamefont {Tong}, \citenamefont {Zheng},\ and\
  \citenamefont {Chen}}]{2010PolyethNanoFibHigThCond}%
  \BibitemOpen
  \bibfield  {author} {\bibinfo {author} {\bibfnamefont {Sheng}\ \bibnamefont
  {Shen}}, \bibinfo {author} {\bibfnamefont {Asegun}\ \bibnamefont {Henry}},
  \bibinfo {author} {\bibfnamefont {Jonathan}\ \bibnamefont {Tong}}, \bibinfo
  {author} {\bibfnamefont {Ruiting}\ \bibnamefont {Zheng}}, \ and\ \bibinfo
  {author} {\bibfnamefont {Gang}\ \bibnamefont {Chen}},\ }\bibfield  {title}
  {\enquote {\bibinfo {title} {Polyethylene nanofibres with very high thermal
  conductivities},}\ }\href {\doibase 10.1038/nnano.2010.27} {\bibfield
  {journal} {\bibinfo  {journal} {Nature Nanotechnology}\ }\textbf {\bibinfo
  {volume} {5}},\ \bibinfo {pages} {251--255} (\bibinfo {year}
  {2010})}\BibitemShut {NoStop}%
\bibitem [{\citenamefont {Gotsmann}\ \emph {et~al.}(2022)\citenamefont
  {Gotsmann}, \citenamefont {Gemma},\ and\ \citenamefont
  {Segal}}]{2022QuantPhTrDvira}%
  \BibitemOpen
  \bibfield  {author} {\bibinfo {author} {\bibfnamefont {Bernd}\ \bibnamefont
  {Gotsmann}}, \bibinfo {author} {\bibfnamefont {Andrea}\ \bibnamefont
  {Gemma}}, \ and\ \bibinfo {author} {\bibfnamefont {Dvira}\ \bibnamefont
  {Segal}},\ }\bibfield  {title} {\enquote {\bibinfo {title} {Quantum phonon
  transport through channels and molecules. a perspective},}\ }\href {\doibase
  10.1063/5.0088460} {\bibfield  {journal} {\bibinfo  {journal} {Applied
  Physics Letters}\ }\textbf {\bibinfo {volume} {120}},\ \bibinfo {pages}
  {160503} (\bibinfo {year} {2022})},\ \Eprint
  {http://arxiv.org/abs/https://pubs.aip.org/aip/apl/article-pdf/doi/10.1063/5.0088460/20034951/160503\_1\_5.0088460.pdf}
  {https://pubs.aip.org/aip/apl/article-pdf/doi/10.1063/5.0088460/20034951/160503\_1\_5.0088460.pdf}
  \BibitemShut {NoStop}%
\bibitem [{\citenamefont {Nitzan}\ and\ \citenamefont
  {Ratner}(2003)}]{NitzanScience03}%
  \BibitemOpen
  \bibfield  {author} {\bibinfo {author} {\bibfnamefont {Abraham}\ \bibnamefont
  {Nitzan}}\ and\ \bibinfo {author} {\bibfnamefont {Mark~A.}\ \bibnamefont
  {Ratner}},\ }\bibfield  {title} {\enquote {\bibinfo {title} {Electron
  transport in molecular wire junctions},}\ }\href {\doibase
  10.1126/science.1081572} {\bibfield  {journal} {\bibinfo  {journal}
  {Science}\ }\textbf {\bibinfo {volume} {300}},\ \bibinfo {pages} {1384--1389}
  (\bibinfo {year} {2003})},\ \Eprint
  {http://arxiv.org/abs/http://science.sciencemag.org/content/300/5624/1384.full.pdf}
  {http://science.sciencemag.org/content/300/5624/1384.full.pdf} \BibitemShut
  {NoStop}%
\bibitem [{\citenamefont {Nitzan}(2007)}]{AbeScience07}%
  \BibitemOpen
  \bibfield  {author} {\bibinfo {author} {\bibfnamefont {A.}~\bibnamefont
  {Nitzan}},\ }\bibfield  {title} {\enquote {\bibinfo {title} {Molecules take
  the heat},}\ }\href@noop {} {\bibfield  {journal} {\bibinfo  {journal}
  {Science}\ }\textbf {\bibinfo {volume} {317}},\ \bibinfo {pages} {759--760}
  (\bibinfo {year} {2007})}\BibitemShut {NoStop}%
\bibitem [{\citenamefont {Segal}\ \emph {et~al.}(2003)\citenamefont {Segal},
  \citenamefont {Nitzan},\ and\ \citenamefont {H\"anggi}}]{SegalNitzan03}%
  \BibitemOpen
  \bibfield  {author} {\bibinfo {author} {\bibfnamefont {Dvira}\ \bibnamefont
  {Segal}}, \bibinfo {author} {\bibfnamefont {Abraham}\ \bibnamefont {Nitzan}},
  \ and\ \bibinfo {author} {\bibfnamefont {Peter}\ \bibnamefont {H\"anggi}},\
  }\bibfield  {title} {\enquote {\bibinfo {title} {Thermal conductance through
  molecular wires},}\ }\href {\doibase 10.1063/1.1603211} {\bibfield  {journal}
  {\bibinfo  {journal} {The Journal of Chemical Physics}\ }\textbf {\bibinfo
  {volume} {119}},\ \bibinfo {pages} {6840--6855} (\bibinfo {year} {2003})},\
  \Eprint {http://arxiv.org/abs/https://doi.org/10.1063/1.1603211}
  {https://doi.org/10.1063/1.1603211} \BibitemShut {NoStop}%
\bibitem [{\citenamefont {Anufriev}\ \emph {et~al.}(2018)\citenamefont
  {Anufriev}, \citenamefont {Gluchko}, \citenamefont {Volz},\ and\
  \citenamefont {Nomura}}]{2018SiNanowLdepThCond}%
  \BibitemOpen
  \bibfield  {author} {\bibinfo {author} {\bibfnamefont {Roman}\ \bibnamefont
  {Anufriev}}, \bibinfo {author} {\bibfnamefont {Sergei}\ \bibnamefont
  {Gluchko}}, \bibinfo {author} {\bibfnamefont {Sebastian}\ \bibnamefont
  {Volz}}, \ and\ \bibinfo {author} {\bibfnamefont {Masahiro}\ \bibnamefont
  {Nomura}},\ }\bibfield  {title} {\enquote {\bibinfo {title} {Quasi-ballistic
  heat conduction due to l{\'e}vy phonon flights in silicon nanowires},}\
  }\href {\doibase 10.1021/acsnano.8b07597} {\bibfield  {journal} {\bibinfo
  {journal} {ACS Nano}\ }\textbf {\bibinfo {volume} {12}},\ \bibinfo {pages}
  {11928--11935} (\bibinfo {year} {2018})}\BibitemShut {NoStop}%
\bibitem [{\citenamefont {Yang}\ \emph {et~al.}(2021)\citenamefont {Yang},
  \citenamefont {Tao}, \citenamefont {Zhu}, \citenamefont {Akter},
  \citenamefont {Wang}, \citenamefont {Pan}, \citenamefont {Zhao},
  \citenamefont {Zhang}, \citenamefont {Xu}, \citenamefont {Chen},
  \citenamefont {Xu}, \citenamefont {Chen}, \citenamefont {Mao},\ and\
  \citenamefont {Li}}]{2021SuperDiffN033}%
  \BibitemOpen
  \bibfield  {author} {\bibinfo {author} {\bibfnamefont {Lin}\ \bibnamefont
  {Yang}}, \bibinfo {author} {\bibfnamefont {Yi}~\bibnamefont {Tao}}, \bibinfo
  {author} {\bibfnamefont {Yanglin}\ \bibnamefont {Zhu}}, \bibinfo {author}
  {\bibfnamefont {Manira}\ \bibnamefont {Akter}}, \bibinfo {author}
  {\bibfnamefont {Ke}~\bibnamefont {Wang}}, \bibinfo {author} {\bibfnamefont
  {Zhiliang}\ \bibnamefont {Pan}}, \bibinfo {author} {\bibfnamefont {Yang}\
  \bibnamefont {Zhao}}, \bibinfo {author} {\bibfnamefont {Qian}\ \bibnamefont
  {Zhang}}, \bibinfo {author} {\bibfnamefont {Ya-Qiong}\ \bibnamefont {Xu}},
  \bibinfo {author} {\bibfnamefont {Renkun}\ \bibnamefont {Chen}}, \bibinfo
  {author} {\bibfnamefont {Terry~T.}\ \bibnamefont {Xu}}, \bibinfo {author}
  {\bibfnamefont {Yunfei}\ \bibnamefont {Chen}}, \bibinfo {author}
  {\bibfnamefont {Zhiqiang}\ \bibnamefont {Mao}}, \ and\ \bibinfo {author}
  {\bibfnamefont {Deyu}\ \bibnamefont {Li}},\ }\bibfield  {title} {\enquote
  {\bibinfo {title} {Observation of superdiffusive phonon transport in aligned
  atomic chains},}\ }\href {\doibase 10.1038/s41565-021-00884-6} {\bibfield
  {journal} {\bibinfo  {journal} {Nature Nanotechnology}\ }\textbf {\bibinfo
  {volume} {16}},\ \bibinfo {pages} {764--768} (\bibinfo {year}
  {2021})}\BibitemShut {NoStop}%
\bibitem [{\citenamefont {Balandin}\ \emph {et~al.}(2022)\citenamefont
  {Balandin}, \citenamefont {Kargar}, \citenamefont {Salguero},\ and\
  \citenamefont {Lake}}]{VanDerWaalsMaterReview2022}%
  \BibitemOpen
  \bibfield  {author} {\bibinfo {author} {\bibfnamefont {Alexander~A.}\
  \bibnamefont {Balandin}}, \bibinfo {author} {\bibfnamefont {Fariborz}\
  \bibnamefont {Kargar}}, \bibinfo {author} {\bibfnamefont {Tina~T.}\
  \bibnamefont {Salguero}}, \ and\ \bibinfo {author} {\bibfnamefont {Roger~K.}\
  \bibnamefont {Lake}},\ }\bibfield  {title} {\enquote {\bibinfo {title}
  {One-dimensional van der waals quantum materials},}\ }\href {\doibase
  https://doi.org/10.1016/j.mattod.2022.03.015} {\bibfield  {journal} {\bibinfo
   {journal} {Materials Today}\ }\textbf {\bibinfo {volume} {55}},\ \bibinfo
  {pages} {74--91} (\bibinfo {year} {2022})}\BibitemShut {NoStop}%
\bibitem [{\citenamefont {Pan}\ \emph {et~al.}(2022)\citenamefont {Pan},
  \citenamefont {Lee}, \citenamefont {Wang}, \citenamefont {Mao},\ and\
  \citenamefont {Li}}]{20221DPhTaSe}%
  \BibitemOpen
  \bibfield  {author} {\bibinfo {author} {\bibfnamefont {Zhiliang}\
  \bibnamefont {Pan}}, \bibinfo {author} {\bibfnamefont {Seng~Huat}\
  \bibnamefont {Lee}}, \bibinfo {author} {\bibfnamefont {Ke}~\bibnamefont
  {Wang}}, \bibinfo {author} {\bibfnamefont {Zhiqiang}\ \bibnamefont {Mao}}, \
  and\ \bibinfo {author} {\bibfnamefont {Deyu}\ \bibnamefont {Li}},\ }\bibfield
   {title} {\enquote {\bibinfo {title} {Elastic stiffening induces
  one-dimensional phonons in thin ta2se3 nanowires},}\ }\href {\doibase
  10.1063/5.0083980} {\bibfield  {journal} {\bibinfo  {journal} {Applied
  Physics Letters}\ }\textbf {\bibinfo {volume} {120}},\ \bibinfo {pages}
  {062201} (\bibinfo {year} {2022})},\ \Eprint
  {http://arxiv.org/abs/https://pubs.aip.org/aip/apl/article-pdf/doi/10.1063/5.0083980/16476981/062201\_1\_online.pdf}
  {https://pubs.aip.org/aip/apl/article-pdf/doi/10.1063/5.0083980/16476981/062201\_1\_online.pdf}
  \BibitemShut {NoStop}%
\bibitem [{\citenamefont {Liu}\ \emph {et~al.}(2023)\citenamefont {Liu},
  \citenamefont {Wu}, \citenamefont {Tan}, \citenamefont {Tao}, \citenamefont
  {Zhang}, \citenamefont {Li}, \citenamefont {Yang}, \citenamefont {Yan},\ and\
  \citenamefont {Chen}}]{2023DopingEnhPhTransp}%
  \BibitemOpen
  \bibfield  {author} {\bibinfo {author} {\bibfnamefont {Chenhan}\ \bibnamefont
  {Liu}}, \bibinfo {author} {\bibfnamefont {Chao}\ \bibnamefont {Wu}}, \bibinfo
  {author} {\bibfnamefont {Xian~Yi}\ \bibnamefont {Tan}}, \bibinfo {author}
  {\bibfnamefont {Yi}~\bibnamefont {Tao}}, \bibinfo {author} {\bibfnamefont
  {Yin}\ \bibnamefont {Zhang}}, \bibinfo {author} {\bibfnamefont {Deyu}\
  \bibnamefont {Li}}, \bibinfo {author} {\bibfnamefont {Juekuan}\ \bibnamefont
  {Yang}}, \bibinfo {author} {\bibfnamefont {Qingyu}\ \bibnamefont {Yan}}, \
  and\ \bibinfo {author} {\bibfnamefont {Yunfei}\ \bibnamefont {Chen}},\
  }\bibfield  {title} {\enquote {\bibinfo {title} {Unexpected doping effects on
  phonon transport in quasi-one-dimensional van der waals crystal tis3
  nanoribbons},}\ }\href {\doibase 10.1038/s41467-023-41425-0} {\bibfield
  {journal} {\bibinfo  {journal} {Nature Communications}\ }\textbf {\bibinfo
  {volume} {14}},\ \bibinfo {pages} {5597} (\bibinfo {year}
  {2023})}\BibitemShut {NoStop}%
\bibitem [{\citenamefont {Zhan}\ \emph {et~al.}(2025)\citenamefont {Zhan},
  \citenamefont {He}, \citenamefont {Wang}, \citenamefont {Luo}, \citenamefont
  {Cui},\ and\ \citenamefont {Zheng}}]{2025VanDerWaalsAtChainRevChin}%
  \BibitemOpen
  \bibfield  {author} {\bibinfo {author} {\bibfnamefont {Pengxin}\ \bibnamefont
  {Zhan}}, \bibinfo {author} {\bibfnamefont {Ping}\ \bibnamefont {He}},
  \bibinfo {author} {\bibfnamefont {Zike}\ \bibnamefont {Wang}}, \bibinfo
  {author} {\bibfnamefont {Lingxin}\ \bibnamefont {Luo}}, \bibinfo {author}
  {\bibfnamefont {Xueping}\ \bibnamefont {Cui}}, \ and\ \bibinfo {author}
  {\bibfnamefont {Jian}\ \bibnamefont {Zheng}},\ }\bibfield  {title} {\enquote
  {\bibinfo {title} {Research progress of one-dimensional van der waals atomic
  chain materials},}\ }\href {\doibase 10.1007/s40843-024-3205-7} {\bibfield
  {journal} {\bibinfo  {journal} {Science China Materials}\ }\textbf {\bibinfo
  {volume} {68}},\ \bibinfo {pages} {364--386} (\bibinfo {year}
  {2025})}\BibitemShut {NoStop}%
\bibitem [{\citenamefont {Rego}\ and\ \citenamefont
  {Kirczenow}(1998)}]{1998QuantThermCond}%
  \BibitemOpen
  \bibfield  {author} {\bibinfo {author} {\bibfnamefont {Luis G.~C.}\
  \bibnamefont {Rego}}\ and\ \bibinfo {author} {\bibfnamefont {George}\
  \bibnamefont {Kirczenow}},\ }\bibfield  {title} {\enquote {\bibinfo {title}
  {Quantized thermal conductance of dielectric quantum wires},}\ }\href
  {\doibase 10.1103/PhysRevLett.81.232} {\bibfield  {journal} {\bibinfo
  {journal} {Phys. Rev. Lett.}\ }\textbf {\bibinfo {volume} {81}},\ \bibinfo
  {pages} {232--235} (\bibinfo {year} {1998})}\BibitemShut {NoStop}%
\bibitem [{\citenamefont {Angelescu}\ \emph {et~al.}(1998)\citenamefont
  {Angelescu}, \citenamefont {Cross},\ and\ \citenamefont
  {Roukes}}]{1998HeatTranspRev}%
  \BibitemOpen
  \bibfield  {author} {\bibinfo {author} {\bibfnamefont {D.E.}\ \bibnamefont
  {Angelescu}}, \bibinfo {author} {\bibfnamefont {M.C.}\ \bibnamefont {Cross}},
  \ and\ \bibinfo {author} {\bibfnamefont {M.L.}\ \bibnamefont {Roukes}},\
  }\bibfield  {title} {\enquote {\bibinfo {title} {Heat transport in mesoscopic
  systems},}\ }\href {\doibase https://doi.org/10.1006/spmi.1997.0561}
  {\bibfield  {journal} {\bibinfo  {journal} {Superlattices and
  Microstructures}\ }\textbf {\bibinfo {volume} {23}},\ \bibinfo {pages}
  {673--689} (\bibinfo {year} {1998})}\BibitemShut {NoStop}%
\bibitem [{\citenamefont {Narayan}\ and\ \citenamefont
  {Ramaswamy}(2002)}]{Turbulence02}%
  \BibitemOpen
  \bibfield  {author} {\bibinfo {author} {\bibfnamefont {Onuttom}\ \bibnamefont
  {Narayan}}\ and\ \bibinfo {author} {\bibfnamefont {Sriram}\ \bibnamefont
  {Ramaswamy}},\ }\bibfield  {title} {\enquote {\bibinfo {title} {Anomalous
  heat conduction in one-dimensional momentum-conserving systems},}\ }\href
  {\doibase 10.1103/PhysRevLett.89.200601} {\bibfield  {journal} {\bibinfo
  {journal} {Phys. Rev. Lett.}\ }\textbf {\bibinfo {volume} {89}},\ \bibinfo
  {pages} {200601} (\bibinfo {year} {2002})}\BibitemShut {NoStop}%
\bibitem [{\citenamefont {Wang}\ \emph {et~al.}(2015)\citenamefont {Wang},
  \citenamefont {Wu},\ and\ \citenamefont {Xu}}]{2015SuperHeatDiffMomCons}%
  \BibitemOpen
  \bibfield  {author} {\bibinfo {author} {\bibfnamefont {Lei}\ \bibnamefont
  {Wang}}, \bibinfo {author} {\bibfnamefont {Zhiyuan}\ \bibnamefont {Wu}}, \
  and\ \bibinfo {author} {\bibfnamefont {Lubo}\ \bibnamefont {Xu}},\ }\bibfield
   {title} {\enquote {\bibinfo {title} {Super heat diffusion in one-dimensional
  momentum-conserving nonlinear lattices},}\ }\href {\doibase
  10.1103/PhysRevE.91.062130} {\bibfield  {journal} {\bibinfo  {journal} {Phys.
  Rev. E}\ }\textbf {\bibinfo {volume} {91}},\ \bibinfo {pages} {062130}
  (\bibinfo {year} {2015})}\BibitemShut {NoStop}%
\bibitem [{\citenamefont {Maire}\ \emph {et~al.}(2017)\citenamefont {Maire},
  \citenamefont {Anufriev},\ and\ \citenamefont
  {Nomura}}]{2017BallTr4KSiNanow033}%
  \BibitemOpen
  \bibfield  {author} {\bibinfo {author} {\bibfnamefont {Jeremie}\ \bibnamefont
  {Maire}}, \bibinfo {author} {\bibfnamefont {Roman}\ \bibnamefont {Anufriev}},
  \ and\ \bibinfo {author} {\bibfnamefont {Masahiro}\ \bibnamefont {Nomura}},\
  }\bibfield  {title} {\enquote {\bibinfo {title} {Ballistic thermal transport
  in silicon nanowires},}\ }\href {\doibase 10.1038/srep41794} {\bibfield
  {journal} {\bibinfo  {journal} {Scientific Reports}\ }\textbf {\bibinfo
  {volume} {7}},\ \bibinfo {pages} {41794} (\bibinfo {year}
  {2017})}\BibitemShut {NoStop}%
\bibitem [{\citenamefont {Liu}\ \emph {et~al.}(2024)\citenamefont {Liu},
  \citenamefont {Jhalaria}, \citenamefont {Ruzicka}, \citenamefont
  {Benicewicz}, \citenamefont {Kumar}, \citenamefont {Fytas},\ and\
  \citenamefont {Xu}}]{2024prlnanopartonpolymsuperdiff}%
  \BibitemOpen
  \bibfield  {author} {\bibinfo {author} {\bibfnamefont {Bohai}\ \bibnamefont
  {Liu}}, \bibinfo {author} {\bibfnamefont {Mayank}\ \bibnamefont {Jhalaria}},
  \bibinfo {author} {\bibfnamefont {Eric}\ \bibnamefont {Ruzicka}}, \bibinfo
  {author} {\bibfnamefont {Brian~C.}\ \bibnamefont {Benicewicz}}, \bibinfo
  {author} {\bibfnamefont {Sanat~K.}\ \bibnamefont {Kumar}}, \bibinfo {author}
  {\bibfnamefont {George}\ \bibnamefont {Fytas}}, \ and\ \bibinfo {author}
  {\bibfnamefont {Xiangfan}\ \bibnamefont {Xu}},\ }\bibfield  {title} {\enquote
  {\bibinfo {title} {Superdiffusive thermal transport in polymer-grafted
  nanoparticle melts},}\ }\href {\doibase 10.1103/PhysRevLett.133.248101}
  {\bibfield  {journal} {\bibinfo  {journal} {Phys. Rev. Lett.}\ }\textbf
  {\bibinfo {volume} {133}},\ \bibinfo {pages} {248101} (\bibinfo {year}
  {2024})}\BibitemShut {NoStop}%
\bibitem [{\citenamefont {Lepri}\ \emph {et~al.}(2003)\citenamefont {Lepri},
  \citenamefont {Livi},\ and\ \citenamefont {Politi}}]{2003RevClassLowD}%
  \BibitemOpen
  \bibfield  {author} {\bibinfo {author} {\bibfnamefont {Stefano}\ \bibnamefont
  {Lepri}}, \bibinfo {author} {\bibfnamefont {Roberto}\ \bibnamefont {Livi}}, \
  and\ \bibinfo {author} {\bibfnamefont {Antonio}\ \bibnamefont {Politi}},\
  }\bibfield  {title} {\enquote {\bibinfo {title} {Thermal conduction in
  classical low-dimensional lattices},}\ }\href {\doibase
  https://doi.org/10.1016/S0370-1573(02)00558-6} {\bibfield  {journal}
  {\bibinfo  {journal} {Physics Reports}\ }\textbf {\bibinfo {volume} {377}},\
  \bibinfo {pages} {1--80} (\bibinfo {year} {2003})}\BibitemShut {NoStop}%
\bibitem [{\citenamefont {Livi}(2023)}]{2021RevLiviPedagog}%
  \BibitemOpen
  \bibfield  {author} {\bibinfo {author} {\bibfnamefont {Roberto}\ \bibnamefont
  {Livi}},\ }\bibfield  {title} {\enquote {\bibinfo {title} {Anomalous
  transport in low-dimensional systems: A pedagogical overview},}\ }\href
  {\doibase https://doi.org/10.1016/j.physa.2022.127779} {\bibfield  {journal}
  {\bibinfo  {journal} {Physica A: Statistical Mechanics and its Applications}\
  }\textbf {\bibinfo {volume} {631}},\ \bibinfo {pages} {127779} (\bibinfo
  {year} {2023})},\ \bibinfo {note} {lecture Notes of the 15th International
  Summer School of Fundamental Problems in Statistical Physics}\BibitemShut
  {NoStop}%
\bibitem [{\citenamefont {Anufriev}\ \emph {et~al.}(2021)\citenamefont
  {Anufriev}, \citenamefont {Wu},\ and\ \citenamefont {Nomura}}]{2021ballist}%
  \BibitemOpen
  \bibfield  {author} {\bibinfo {author} {\bibfnamefont {Roman}\ \bibnamefont
  {Anufriev}}, \bibinfo {author} {\bibfnamefont {Yunhui}\ \bibnamefont {Wu}}, \
  and\ \bibinfo {author} {\bibfnamefont {Masahiro}\ \bibnamefont {Nomura}},\
  }\bibfield  {title} {\enquote {\bibinfo {title} {Ballistic heat conduction in
  semiconductor nanowires},}\ }\href {\doibase 10.1063/5.0060026} {\bibfield
  {journal} {\bibinfo  {journal} {Journal of Applied Physics}\ }\textbf
  {\bibinfo {volume} {130}},\ \bibinfo {pages} {070903} (\bibinfo {year}
  {2021})},\ \Eprint
  {http://arxiv.org/abs/https://pubs.aip.org/aip/jap/article-pdf/doi/10.1063/5.0060026/20029443/070903\_1\_5.0060026.pdf}
  {https://pubs.aip.org/aip/jap/article-pdf/doi/10.1063/5.0060026/20029443/070903\_1\_5.0060026.pdf}
  \BibitemShut {NoStop}%
\bibitem [{\citenamefont {Schwab}\ \emph {et~al.}(2000)\citenamefont {Schwab},
  \citenamefont {Henriksen}, \citenamefont {Worlock},\ and\ \citenamefont
  {Roukes}}]{2000NatQuantThermCond}%
  \BibitemOpen
  \bibfield  {author} {\bibinfo {author} {\bibfnamefont {K.}~\bibnamefont
  {Schwab}}, \bibinfo {author} {\bibfnamefont {E.~A.}\ \bibnamefont
  {Henriksen}}, \bibinfo {author} {\bibfnamefont {J.~M.}\ \bibnamefont
  {Worlock}}, \ and\ \bibinfo {author} {\bibfnamefont {M.~L.}\ \bibnamefont
  {Roukes}},\ }\bibfield  {title} {\enquote {\bibinfo {title} {Measurement of
  the quantum of thermal conductance},}\ }\href {\doibase 10.1038/35010065}
  {\bibfield  {journal} {\bibinfo  {journal} {Nature}\ }\textbf {\bibinfo
  {volume} {404}},\ \bibinfo {pages} {974--977} (\bibinfo {year}
  {2000})}\BibitemShut {NoStop}%
\bibitem [{\citenamefont {Rubtsova}\ \emph {et~al.}(2015)\citenamefont
  {Rubtsova}, \citenamefont {Nyby}, \citenamefont {Zhang}, \citenamefont
  {Zhang}, \citenamefont {Zhou}, \citenamefont {Jayawickramarajah},
  \citenamefont {Burin},\ and\ \citenamefont
  {Rubtsov}}]{ab15ballistictranspexp}%
  \BibitemOpen
  \bibfield  {author} {\bibinfo {author} {\bibfnamefont {Natalia~I.}\
  \bibnamefont {Rubtsova}}, \bibinfo {author} {\bibfnamefont {Clara~M.}\
  \bibnamefont {Nyby}}, \bibinfo {author} {\bibfnamefont {Hong}\ \bibnamefont
  {Zhang}}, \bibinfo {author} {\bibfnamefont {Boyu}\ \bibnamefont {Zhang}},
  \bibinfo {author} {\bibfnamefont {Xiao}\ \bibnamefont {Zhou}}, \bibinfo
  {author} {\bibfnamefont {Janarthanan}\ \bibnamefont {Jayawickramarajah}},
  \bibinfo {author} {\bibfnamefont {Alexander~L.}\ \bibnamefont {Burin}}, \
  and\ \bibinfo {author} {\bibfnamefont {Igor~V.}\ \bibnamefont {Rubtsov}},\
  }\bibfield  {title} {\enquote {\bibinfo {title} {Room-temperature ballistic
  energy transport in molecules with repeating units},}\ }\href {\doibase
  10.1063/1.4916326} {\bibfield  {journal} {\bibinfo  {journal} {The Journal of
  Chemical Physics}\ }\textbf {\bibinfo {volume} {142}},\ \bibinfo {pages}
  {212412} (\bibinfo {year} {2015})},\ \Eprint
  {http://arxiv.org/abs/https://doi.org/10.1063/1.4916326}
  {https://doi.org/10.1063/1.4916326} \BibitemShut {NoStop}%
\bibitem [{\citenamefont {Rubtsov}\ and\ \citenamefont
  {Burin}(2019)}]{ab19IgorReview}%
  \BibitemOpen
  \bibfield  {author} {\bibinfo {author} {\bibfnamefont {Igor~V.}\ \bibnamefont
  {Rubtsov}}\ and\ \bibinfo {author} {\bibfnamefont {Alexander~L.}\
  \bibnamefont {Burin}},\ }\bibfield  {title} {\enquote {\bibinfo {title}
  {Ballistic and diffusive vibrational energy transport in molecules},}\ }\href
  {\doibase 10.1063/1.5055670} {\bibfield  {journal} {\bibinfo  {journal} {The
  Journal of Chemical Physics}\ }\textbf {\bibinfo {volume} {150}},\ \bibinfo
  {pages} {020901} (\bibinfo {year} {2019})},\ \Eprint
  {http://arxiv.org/abs/https://doi.org/10.1063/1.5055670}
  {https://doi.org/10.1063/1.5055670} \BibitemShut {NoStop}%
\bibitem [{\citenamefont {Fermi}\ \emph {et~al.}(1955)\citenamefont {Fermi},
  \citenamefont {Pasta}, \citenamefont {Ulam},\ and\ \citenamefont
  {Tsingou}}]{FPUclassic}%
  \BibitemOpen
  \bibfield  {author} {\bibinfo {author} {\bibfnamefont {E.}~\bibnamefont
  {Fermi}}, \bibinfo {author} {\bibfnamefont {J.R.}\ \bibnamefont {Pasta}},
  \bibinfo {author} {\bibfnamefont {S.}~\bibnamefont {Ulam}}, \ and\ \bibinfo
  {author} {\bibfnamefont {M.}~\bibnamefont {Tsingou}},\ }\bibfield  {title}
  {\enquote {\bibinfo {title} {Studies of the nonlinear problems},}\
  }\href@noop {} {\bibfield  {journal} {\bibinfo  {journal} {Los Alamos
  Scientific Laboratory of the University of Califormia}\ } (\bibinfo {year}
  {1955})}\BibitemShut {NoStop}%
\bibitem [{\citenamefont {Wang}\ and\ \citenamefont {Li}(2004)}]{2004TrVibrMD}%
  \BibitemOpen
  \bibfield  {author} {\bibinfo {author} {\bibfnamefont {Jian-Sheng}\
  \bibnamefont {Wang}}\ and\ \bibinfo {author} {\bibfnamefont {Baowen}\
  \bibnamefont {Li}},\ }\bibfield  {title} {\enquote {\bibinfo {title}
  {Mode-coupling theory and molecular dynamics simulation for heat conduction
  in a chain with transverse motions},}\ }\href {\doibase
  10.1103/PhysRevE.70.021204} {\bibfield  {journal} {\bibinfo  {journal} {Phys.
  Rev. E}\ }\textbf {\bibinfo {volume} {70}},\ \bibinfo {pages} {021204}
  (\bibinfo {year} {2004})}\BibitemShut {NoStop}%
\bibitem [{\citenamefont {Mendl}\ and\ \citenamefont
  {Spohn}(2013)}]{2013FPUChainExactMendl}%
  \BibitemOpen
  \bibfield  {author} {\bibinfo {author} {\bibfnamefont {Christian~B.}\
  \bibnamefont {Mendl}}\ and\ \bibinfo {author} {\bibfnamefont {Herbert}\
  \bibnamefont {Spohn}},\ }\bibfield  {title} {\enquote {\bibinfo {title}
  {Dynamic correlators of fermi-pasta-ulam chains and nonlinear fluctuating
  hydrodynamics},}\ }\href {\doibase 10.1103/PhysRevLett.111.230601} {\bibfield
   {journal} {\bibinfo  {journal} {Phys. Rev. Lett.}\ }\textbf {\bibinfo
  {volume} {111}},\ \bibinfo {pages} {230601} (\bibinfo {year}
  {2013})}\BibitemShut {NoStop}%
\bibitem [{\citenamefont {Swinteck}\ \emph {et~al.}(2013)\citenamefont
  {Swinteck}, \citenamefont {Muralidharan},\ and\ \citenamefont
  {Deymier}}]{2013FPUTAnalNum}%
  \BibitemOpen
  \bibfield  {author} {\bibinfo {author} {\bibfnamefont {Nichlas~Z.}\
  \bibnamefont {Swinteck}}, \bibinfo {author} {\bibfnamefont {Krishna}\
  \bibnamefont {Muralidharan}}, \ and\ \bibinfo {author} {\bibfnamefont
  {Pierre~A.}\ \bibnamefont {Deymier}},\ }\bibfield  {title} {\enquote
  {\bibinfo {title} {Phonon scattering in one-dimensional anharmonic crystals
  and superlattices: Analytical and numerical study},}\ }\href {\doibase
  10.1115/1.4023824} {\bibfield  {journal} {\bibinfo  {journal} {Journal of
  Vibration and Acoustics}\ }\textbf {\bibinfo {volume} {135}},\ \bibinfo
  {pages} {041016} (\bibinfo {year} {2013})},\ \Eprint
  {http://arxiv.org/abs/https://asmedigitalcollection.asme.org/vibrationacoustics/article-pdf/135/4/041016/6339365/vib\_135\_4\_041016.pdf}
  {https://asmedigitalcollection.asme.org/vibrationacoustics/article-pdf/135/4/041016/6339365/vib\_135\_4\_041016.pdf}
  \BibitemShut {NoStop}%
\bibitem [{\citenamefont {Rieder}\ \emph {et~al.}(1967)\citenamefont {Rieder},
  \citenamefont {Lebowitz},\ and\ \citenamefont {Lieb}}]{Lebowitz67}%
  \BibitemOpen
  \bibfield  {author} {\bibinfo {author} {\bibfnamefont {Z.}~\bibnamefont
  {Rieder}}, \bibinfo {author} {\bibfnamefont {J.~L.}\ \bibnamefont
  {Lebowitz}}, \ and\ \bibinfo {author} {\bibfnamefont {E.}~\bibnamefont
  {Lieb}},\ }\bibfield  {title} {\enquote {\bibinfo {title} {Properties of a
  harmonic crystal in a stationary nonequilibrium state},}\ }\href {\doibase
  10.1063/1.1705319} {\bibfield  {journal} {\bibinfo  {journal} {Journal of
  Mathematical Physics}\ }\textbf {\bibinfo {volume} {8}},\ \bibinfo {pages}
  {1073--1078} (\bibinfo {year} {1967})},\ \Eprint
  {http://arxiv.org/abs/https://doi.org/10.1063/1.1705319}
  {https://doi.org/10.1063/1.1705319} \BibitemShut {NoStop}%
\bibitem [{\citenamefont {Bonetto}\ \emph {et~al.}(2000)\citenamefont
  {Bonetto}, \citenamefont {Lebowitz},\ and\ \citenamefont
  {Rey-Bellet}}]{LEBOWITZ00}%
  \BibitemOpen
  \bibfield  {author} {\bibinfo {author} {\bibfnamefont {F.}~\bibnamefont
  {Bonetto}}, \bibinfo {author} {\bibfnamefont {J.~L.}\ \bibnamefont
  {Lebowitz}}, \ and\ \bibinfo {author} {\bibfnamefont {L.}~\bibnamefont
  {Rey-Bellet}}\ }(\bibinfo  {publisher} {Published by Imperial College Press,
  distributed by World Scientific Publishing CO.},\ \bibinfo {year} {2000})\
  Chap.\ \bibinfo {chapter} {Fourier's Law: a Challenge to Theorists}, pp.\
  \bibinfo {pages} {128--150},\ \bibinfo {note} {0}\BibitemShut {NoStop}%
\bibitem [{\citenamefont {Leitner}(2001)}]{20011DGlass}%
  \BibitemOpen
  \bibfield  {author} {\bibinfo {author} {\bibfnamefont {David~M.}\
  \bibnamefont {Leitner}},\ }\bibfield  {title} {\enquote {\bibinfo {title}
  {Vibrational energy transfer and heat conduction in a one-dimensional
  glass},}\ }\href {\doibase 10.1103/PhysRevB.64.094201} {\bibfield  {journal}
  {\bibinfo  {journal} {Phys. Rev. B}\ }\textbf {\bibinfo {volume} {64}},\
  \bibinfo {pages} {094201} (\bibinfo {year} {2001})}\BibitemShut {NoStop}%
\bibitem [{\citenamefont {Landauer}(1957)}]{Landauer57Classic}%
  \BibitemOpen
  \bibfield  {author} {\bibinfo {author} {\bibfnamefont {R.}~\bibnamefont
  {Landauer}},\ }\bibfield  {title} {\enquote {\bibinfo {title} {Spatial
  variation of currents and fields due to localized scatterers in metallic
  conduction},}\ }\href {\doibase 10.1147/rd.13.0223} {\bibfield  {journal}
  {\bibinfo  {journal} {IBM Journal of Research and Development}\ }\textbf
  {\bibinfo {volume} {1}},\ \bibinfo {pages} {223--231} (\bibinfo {year}
  {1957})}\BibitemShut {NoStop}%
\bibitem [{\citenamefont {Klitsner}\ \emph {et~al.}(1988)\citenamefont
  {Klitsner}, \citenamefont {VanCleve}, \citenamefont {Fischer},\ and\
  \citenamefont {Pohl}}]{1988EquatForThermCond}%
  \BibitemOpen
  \bibfield  {author} {\bibinfo {author} {\bibfnamefont {Tom}\ \bibnamefont
  {Klitsner}}, \bibinfo {author} {\bibfnamefont {J.~E.}\ \bibnamefont
  {VanCleve}}, \bibinfo {author} {\bibfnamefont {Henry~E.}\ \bibnamefont
  {Fischer}}, \ and\ \bibinfo {author} {\bibfnamefont {R.~O.}\ \bibnamefont
  {Pohl}},\ }\bibfield  {title} {\enquote {\bibinfo {title} {Phonon radiative
  heat transfer and surface scattering},}\ }\href {\doibase
  10.1103/PhysRevB.38.7576} {\bibfield  {journal} {\bibinfo  {journal} {Phys.
  Rev. B}\ }\textbf {\bibinfo {volume} {38}},\ \bibinfo {pages} {7576--7594}
  (\bibinfo {year} {1988})}\BibitemShut {NoStop}%
\bibitem [{\citenamefont {Bird}\ \emph {et~al.}(2002)\citenamefont {Bird},
  \citenamefont {Stewart},\ and\ \citenamefont
  {Lightfoot}}]{TransportTextbird2002transport}%
  \BibitemOpen
  \bibfield  {author} {\bibinfo {author} {\bibfnamefont {R.B.}\ \bibnamefont
  {Bird}}, \bibinfo {author} {\bibfnamefont {W.E.}\ \bibnamefont {Stewart}}, \
  and\ \bibinfo {author} {\bibfnamefont {E.N.}\ \bibnamefont {Lightfoot}},\
  }\href {https://books.google.com/books?id=wYnRQwAACAAJ} {\emph {\bibinfo
  {title} {Transport Phenomena}}}\ (\bibinfo  {publisher} {J. Wiley},\ \bibinfo
  {year} {2002})\BibitemShut {NoStop}%
\bibitem [{\citenamefont {Lepri}(1998)}]{1998LepriAnomalThCondFPUCl}%
  \BibitemOpen
  \bibfield  {author} {\bibinfo {author} {\bibfnamefont {Stefano}\ \bibnamefont
  {Lepri}},\ }\bibfield  {title} {\enquote {\bibinfo {title} {Relaxation of
  classical many-body hamiltonians in one dimension},}\ }\href {\doibase
  10.1103/PhysRevE.58.7165} {\bibfield  {journal} {\bibinfo  {journal} {Phys.
  Rev. E}\ }\textbf {\bibinfo {volume} {58}},\ \bibinfo {pages} {7165--7171}
  (\bibinfo {year} {1998})}\BibitemShut {NoStop}%
\bibitem [{\citenamefont {Sharony}\ \emph {et~al.}(2020)\citenamefont
  {Sharony}, \citenamefont {Chen},\ and\ \citenamefont
  {Nitzan}}]{2020MDLangEnergyTrNitzan}%
  \BibitemOpen
  \bibfield  {author} {\bibinfo {author} {\bibfnamefont {Inon}\ \bibnamefont
  {Sharony}}, \bibinfo {author} {\bibfnamefont {Renai}\ \bibnamefont {Chen}}, \
  and\ \bibinfo {author} {\bibfnamefont {Abraham}\ \bibnamefont {Nitzan}},\
  }\bibfield  {title} {\enquote {\bibinfo {title} {Stochastic simulation of
  nonequilibrium heat conduction in extended molecular junctions},}\ }\href
  {\doibase 10.1063/5.0022423} {\bibfield  {journal} {\bibinfo  {journal} {The
  Journal of Chemical Physics}\ }\textbf {\bibinfo {volume} {153}},\ \bibinfo
  {pages} {144113} (\bibinfo {year} {2020})},\ \Eprint
  {http://arxiv.org/abs/https://pubs.aip.org/aip/jcp/article-pdf/doi/10.1063/5.0022423/13370844/144113\_1\_online.pdf}
  {https://pubs.aip.org/aip/jcp/article-pdf/doi/10.1063/5.0022423/13370844/144113\_1\_online.pdf}
  \BibitemShut {NoStop}%
\bibitem [{\citenamefont {Liang}\ \emph {et~al.}(2022)\citenamefont {Liang},
  \citenamefont {Xu}, \citenamefont {Han}, \citenamefont {Yao}, \citenamefont
  {Zhang}, \citenamefont {Zeng}, \citenamefont {Xu},\ and\ \citenamefont
  {Wu}}]{2022highthcond-diam-nanow}%
  \BibitemOpen
  \bibfield  {author} {\bibinfo {author} {\bibfnamefont {T.}~\bibnamefont
  {Liang}}, \bibinfo {author} {\bibfnamefont {K.}~\bibnamefont {Xu}}, \bibinfo
  {author} {\bibfnamefont {M.}~\bibnamefont {Han}}, \bibinfo {author}
  {\bibfnamefont {Y.}~\bibnamefont {Yao}}, \bibinfo {author} {\bibfnamefont
  {Z.}~\bibnamefont {Zhang}}, \bibinfo {author} {\bibfnamefont
  {X.}~\bibnamefont {Zeng}}, \bibinfo {author} {\bibfnamefont {J.}~\bibnamefont
  {Xu}}, \ and\ \bibinfo {author} {\bibfnamefont {J.}~\bibnamefont {Wu}},\
  }\bibfield  {title} {\enquote {\bibinfo {title} {Abnormally high thermal
  conductivity in fivefold twinned diamond nanowires},}\ }\href {\doibase
  https://doi.org/10.1016/j.mtphys.2022.100705} {\bibfield  {journal} {\bibinfo
   {journal} {Materials Today Physics}\ }\textbf {\bibinfo {volume} {25}},\
  \bibinfo {pages} {100705} (\bibinfo {year} {2022})}\BibitemShut {NoStop}%
\bibitem [{\citenamefont {Hua}\ \emph {et~al.}(2024)\citenamefont {Hua},
  \citenamefont {Jiang}, \citenamefont {Zhao}, \citenamefont {Shi},
  \citenamefont {Chen}, \citenamefont {Liang}, \citenamefont {Dong},
  \citenamefont {Song},\ and\ \citenamefont
  {Dong}}]{2024SubSuperDiffTranspThRectif}%
  \BibitemOpen
  \bibfield  {author} {\bibinfo {author} {\bibfnamefont {Renjie}\ \bibnamefont
  {Hua}}, \bibinfo {author} {\bibfnamefont {Yunlei}\ \bibnamefont {Jiang}},
  \bibinfo {author} {\bibfnamefont {Zhiguo}\ \bibnamefont {Zhao}}, \bibinfo
  {author} {\bibfnamefont {Lei}\ \bibnamefont {Shi}}, \bibinfo {author}
  {\bibfnamefont {Yiwei}\ \bibnamefont {Chen}}, \bibinfo {author}
  {\bibfnamefont {Suxia}\ \bibnamefont {Liang}}, \bibinfo {author}
  {\bibfnamefont {Ruo-Yu}\ \bibnamefont {Dong}}, \bibinfo {author}
  {\bibfnamefont {Yingru}\ \bibnamefont {Song}}, \ and\ \bibinfo {author}
  {\bibfnamefont {Yuan}\ \bibnamefont {Dong}},\ }\bibfield  {title} {\enquote
  {\bibinfo {title} {Subdiffusive and superdiffusive phonon transport induced
  significant thermal rectification across a one-dimensional kapitza
  interface},}\ }\href {\doibase
  https://doi.org/10.1016/j.ijheatmasstransfer.2024.126113} {\bibfield
  {journal} {\bibinfo  {journal} {International Journal of Heat and Mass
  Transfer}\ }\textbf {\bibinfo {volume} {234}},\ \bibinfo {pages} {126113}
  (\bibinfo {year} {2024})}\BibitemShut {NoStop}%
\bibitem [{\citenamefont {Xu}\ \emph {et~al.}(2023)\citenamefont {Xu},
  \citenamefont {Fan},\ and\ \citenamefont {Zhou}}]{2023MDSimulRev}%
  \BibitemOpen
  \bibfield  {author} {\bibinfo {author} {\bibfnamefont {Yi-Xin}\ \bibnamefont
  {Xu}}, \bibinfo {author} {\bibfnamefont {Hong-Zhao}\ \bibnamefont {Fan}}, \
  and\ \bibinfo {author} {\bibfnamefont {Yan-Guang}\ \bibnamefont {Zhou}},\
  }\bibfield  {title} {\enquote {\bibinfo {title} {Quantifying spectral thermal
  transport properties in framework of molecular dynamics simulations: a
  comprehensive review},}\ }\href {\doibase 10.1007/s12598-023-02483-x}
  {\bibfield  {journal} {\bibinfo  {journal} {Rare Metals}\ }\textbf {\bibinfo
  {volume} {42}},\ \bibinfo {pages} {3914--3944} (\bibinfo {year}
  {2023})}\BibitemShut {NoStop}%
\bibitem [{\citenamefont {Duan}\ \emph {et~al.}(2019)\citenamefont {Duan},
  \citenamefont {Li}, \citenamefont {Liu}, \citenamefont {Chen},\ and\
  \citenamefont {Li}}]{ThermCondKinks2019Xuhui}%
  \BibitemOpen
  \bibfield  {author} {\bibinfo {author} {\bibfnamefont {Xuhui}\ \bibnamefont
  {Duan}}, \bibinfo {author} {\bibfnamefont {Zehuan}\ \bibnamefont {Li}},
  \bibinfo {author} {\bibfnamefont {Jun}\ \bibnamefont {Liu}}, \bibinfo
  {author} {\bibfnamefont {Gang}\ \bibnamefont {Chen}}, \ and\ \bibinfo
  {author} {\bibfnamefont {Xiaobo}\ \bibnamefont {Li}},\ }\bibfield  {title}
  {\enquote {\bibinfo {title} {Roles of kink on the thermal transport in single
  polyethylene chains},}\ }\href {\doibase 10.1063/1.5086453} {\bibfield
  {journal} {\bibinfo  {journal} {Journal of Applied Physics}\ }\textbf
  {\bibinfo {volume} {125}},\ \bibinfo {pages} {164303} (\bibinfo {year}
  {2019})},\ \Eprint
  {http://arxiv.org/abs/https://pubs.aip.org/aip/jap/article-pdf/doi/10.1063/1.5086453/15226427/164303\_1\_online.pdf}
  {https://pubs.aip.org/aip/jap/article-pdf/doi/10.1063/1.5086453/15226427/164303\_1\_online.pdf}
  \BibitemShut {NoStop}%
\bibitem [{\citenamefont {Krivtsov}\ \emph {et~al.}(2024)\citenamefont
  {Krivtsov}, \citenamefont {Kuzkin},\ and\ \citenamefont
  {Tsaplin}}]{2023ChainWithBrks}%
  \BibitemOpen
  \bibfield  {author} {\bibinfo {author} {\bibfnamefont {Anton~M.}\
  \bibnamefont {Krivtsov}}, \bibinfo {author} {\bibfnamefont {Vitaly~A.}\
  \bibnamefont {Kuzkin}}, \ and\ \bibinfo {author} {\bibfnamefont {Vadim~A.}\
  \bibnamefont {Tsaplin}},\ }\bibfield  {title} {\enquote {\bibinfo {title}
  {Transition from ballistic to diffusive heat transfer in a chain with
  breaks},}\ }\href {\doibase 10.1103/PhysRevE.110.054123} {\bibfield
  {journal} {\bibinfo  {journal} {Phys. Rev. E}\ }\textbf {\bibinfo {volume}
  {110}},\ \bibinfo {pages} {054123} (\bibinfo {year} {2024})}\BibitemShut
  {NoStop}%
\bibitem [{\citenamefont {Barbalinardo}\ \emph {et~al.}(2021)\citenamefont
  {Barbalinardo}, \citenamefont {Chen}, \citenamefont {Dong}, \citenamefont
  {Fan},\ and\ \citenamefont
  {Donadio}}]{2021prlCarmonNanotubeSimulHighConvThermCond}%
  \BibitemOpen
  \bibfield  {author} {\bibinfo {author} {\bibfnamefont {Giuseppe}\
  \bibnamefont {Barbalinardo}}, \bibinfo {author} {\bibfnamefont {Zekun}\
  \bibnamefont {Chen}}, \bibinfo {author} {\bibfnamefont {Haikuan}\
  \bibnamefont {Dong}}, \bibinfo {author} {\bibfnamefont {Zheyong}\
  \bibnamefont {Fan}}, \ and\ \bibinfo {author} {\bibfnamefont {Davide}\
  \bibnamefont {Donadio}},\ }\bibfield  {title} {\enquote {\bibinfo {title}
  {Ultrahigh convergent thermal conductivity of carbon nanotubes from
  comprehensive atomistic modeling},}\ }\href {\doibase
  10.1103/PhysRevLett.127.025902} {\bibfield  {journal} {\bibinfo  {journal}
  {Phys. Rev. Lett.}\ }\textbf {\bibinfo {volume} {127}},\ \bibinfo {pages}
  {025902} (\bibinfo {year} {2021})}\BibitemShut {NoStop}%
\bibitem [{\citenamefont {Landau}\ \emph
  {et~al.}(1986{\natexlab{a}})\citenamefont {Landau}, \citenamefont
  {Lif{\v{s}}ic}, \citenamefont {Lifshitz}, \citenamefont {Kosevich},
  \citenamefont {Sykes}, \citenamefont {Pitaevskii},\ and\ \citenamefont
  {Reid}}]{landau1986theory}%
  \BibitemOpen
  \bibfield  {author} {\bibinfo {author} {\bibfnamefont {L.D.}\ \bibnamefont
  {Landau}}, \bibinfo {author} {\bibfnamefont {E.M.}\ \bibnamefont
  {Lif{\v{s}}ic}}, \bibinfo {author} {\bibfnamefont {E.M.}\ \bibnamefont
  {Lifshitz}}, \bibinfo {author} {\bibfnamefont {A.M.}\ \bibnamefont
  {Kosevich}}, \bibinfo {author} {\bibfnamefont {J.B.}\ \bibnamefont {Sykes}},
  \bibinfo {author} {\bibfnamefont {L.P.}\ \bibnamefont {Pitaevskii}}, \ and\
  \bibinfo {author} {\bibfnamefont {W.H.}\ \bibnamefont {Reid}},\ }\href
  {https://books.google.com/books?id=tpY-VkwCkAIC} {\emph {\bibinfo {title}
  {Theory of Elasticity: Volume 7}}},\ Course of theoretical physics\ (\bibinfo
   {publisher} {Elsevier Science},\ \bibinfo {year} {1986})\BibitemShut
  {NoStop}%
\bibitem [{\citenamefont {Hartwig}(1994)}]{PolimVibr1994Book}%
  \BibitemOpen
  \bibfield  {author} {\bibinfo {author} {\bibfnamefont {Günther}\ \bibnamefont
  {Hartwig}},\ }\href {https://doi.org/10.1007/978-1-4757-6213-6} {\emph
  {\bibinfo {title} {Polymer Properties at Room and Cryogenic Temperatures}}}\
  (\bibinfo  {publisher} {Springer, Boston, MA},\ \bibinfo {year} {1994})\ pp.\
  \bibinfo {pages} {17--46}\BibitemShut {NoStop}%
\bibitem [{\citenamefont {Chico}\ \emph {et~al.}(2006)\citenamefont {Chico},
  \citenamefont {P\'erez-\'Alvarez},\ and\ \citenamefont
  {Cabrillo}}]{Chico2006VibrCarbNanot}%
  \BibitemOpen
  \bibfield  {author} {\bibinfo {author} {\bibfnamefont {L.}~\bibnamefont
  {Chico}}, \bibinfo {author} {\bibfnamefont {R.}~\bibnamefont
  {P\'erez-\'Alvarez}}, \ and\ \bibinfo {author} {\bibfnamefont
  {C.}~\bibnamefont {Cabrillo}},\ }\bibfield  {title} {\enquote {\bibinfo
  {title} {Low-frequency phonons in carbon nanotubes: A continuum approach},}\
  }\href {\doibase 10.1103/PhysRevB.73.075425} {\bibfield  {journal} {\bibinfo
  {journal} {Phys. Rev. B}\ }\textbf {\bibinfo {volume} {73}},\ \bibinfo
  {pages} {075425} (\bibinfo {year} {2006})}\BibitemShut {NoStop}%
\bibitem [{\citenamefont {Kuzkin}(2010)}]{2010TransvVibr1D}%
  \BibitemOpen
  \bibfield  {author} {\bibinfo {author} {\bibfnamefont {Vitaly~A.}\
  \bibnamefont {Kuzkin}},\ }\bibfield  {title} {\enquote {\bibinfo {title}
  {Interatomic force in systems with multibody interactions},}\ }\href
  {\doibase 10.1103/PhysRevE.82.016704} {\bibfield  {journal} {\bibinfo
  {journal} {Phys. Rev. E}\ }\textbf {\bibinfo {volume} {82}},\ \bibinfo
  {pages} {016704} (\bibinfo {year} {2010})}\BibitemShut {NoStop}%
\bibitem [{\citenamefont {Boulatov}\ and\ \citenamefont
  {Burin}(2020)}]{ab20Transv}%
  \BibitemOpen
  \bibfield  {author} {\bibinfo {author} {\bibfnamefont {Alexei}\ \bibnamefont
  {Boulatov}}\ and\ \bibinfo {author} {\bibfnamefont {Alexander~L.}\
  \bibnamefont {Burin}},\ }\bibfield  {title} {\enquote {\bibinfo {title}
  {Crucial effect of transverse vibrations on the transport through polymer
  chains},}\ }\href {\doibase 10.1063/5.0018591} {\bibfield  {journal}
  {\bibinfo  {journal} {The Journal of Chemical Physics}\ }\textbf {\bibinfo
  {volume} {153}},\ \bibinfo {pages} {134102} (\bibinfo {year} {2020})},\
  \Eprint {http://arxiv.org/abs/https://doi.org/10.1063/5.0018591}
  {https://doi.org/10.1063/5.0018591} \BibitemShut {NoStop}%
\bibitem [{\citenamefont {Burin}\ \emph {et~al.}(2023)\citenamefont {Burin},
  \citenamefont {Parshin},\ and\ \citenamefont {Rubtsov}}]{ab2023Cherenkov}%
  \BibitemOpen
  \bibfield  {author} {\bibinfo {author} {\bibfnamefont {Alexander~L.}\
  \bibnamefont {Burin}}, \bibinfo {author} {\bibfnamefont {Igor~V.}\
  \bibnamefont {Parshin}}, \ and\ \bibinfo {author} {\bibfnamefont {Igor~V.}\
  \bibnamefont {Rubtsov}},\ }\bibfield  {title} {\enquote {\bibinfo {title}
  {{Maximum propagation speed and Cherenkov effect in optical phonon transport
  through periodic molecular chains}},}\ }\href {\doibase 10.1063/5.0158201}
  {\bibfield  {journal} {\bibinfo  {journal} {The Journal of Chemical Physics}\
  }\textbf {\bibinfo {volume} {159}},\ \bibinfo {pages} {054903} (\bibinfo
  {year} {2023})},\ \Eprint
  {http://arxiv.org/abs/https://pubs.aip.org/aip/jcp/article-pdf/doi/10.1063/5.0158201/18069436/054903\_1\_5.0158201.pdf}
  {https://pubs.aip.org/aip/jcp/article-pdf/doi/10.1063/5.0158201/18069436/054903\_1\_5.0158201.pdf}
  \BibitemShut {NoStop}%
\bibitem [{\citenamefont {Burin}\ and\ \citenamefont
  {Rubtsov}(2024)}]{ab242stagedec}%
  \BibitemOpen
  \bibfield  {author} {\bibinfo {author} {\bibfnamefont {Alexander~L.}\
  \bibnamefont {Burin}}\ and\ \bibinfo {author} {\bibfnamefont {Igor~V.}\
  \bibnamefont {Rubtsov}},\ }\bibfield  {title} {\enquote {\bibinfo {title}
  {Two stage decoherence of optical phonons in long oligomers},}\ }\href
  {\doibase 10.1063/5.0222580} {\bibfield  {journal} {\bibinfo  {journal} {The
  Journal of Chemical Physics}\ }\textbf {\bibinfo {volume} {161}},\ \bibinfo
  {pages} {094901} (\bibinfo {year} {2024})},\ \Eprint
  {http://arxiv.org/abs/https://pubs.aip.org/aip/jcp/article-pdf/doi/10.1063/5.0222580/20138643/094901\_1\_5.0222580.pdf}
  {https://pubs.aip.org/aip/jcp/article-pdf/doi/10.1063/5.0222580/20138643/094901\_1\_5.0222580.pdf}
  \BibitemShut {NoStop}%
\bibitem [{\citenamefont {Schick}(1968)}]{SchickFPULuttLiq68}%
  \BibitemOpen
  \bibfield  {author} {\bibinfo {author} {\bibfnamefont {Michael}\ \bibnamefont
  {Schick}},\ }\bibfield  {title} {\enquote {\bibinfo {title} {Flux
  quantization in a one-dimensional model},}\ }\href {\doibase
  10.1103/PhysRev.166.404} {\bibfield  {journal} {\bibinfo  {journal} {Phys.
  Rev.}\ }\textbf {\bibinfo {volume} {166}},\ \bibinfo {pages} {404--414}
  (\bibinfo {year} {1968})}\BibitemShut {NoStop}%
\bibitem [{\citenamefont {Okamoto}\ \emph {et~al.}(2019)\citenamefont
  {Okamoto}, \citenamefont {Yanagisawa}, \citenamefont {Anufriev},
  \citenamefont {Mahfuz~Alam}, \citenamefont {Sawano}, \citenamefont
  {Kurosawa},\ and\ \citenamefont {Nomura}}]{2019IncrThCondSiGe}%
  \BibitemOpen
  \bibfield  {author} {\bibinfo {author} {\bibfnamefont {Noboru}\ \bibnamefont
  {Okamoto}}, \bibinfo {author} {\bibfnamefont {Ryoto}\ \bibnamefont
  {Yanagisawa}}, \bibinfo {author} {\bibfnamefont {Roman}\ \bibnamefont
  {Anufriev}}, \bibinfo {author} {\bibfnamefont {Md.}\ \bibnamefont
  {Mahfuz~Alam}}, \bibinfo {author} {\bibfnamefont {Kentarou}\ \bibnamefont
  {Sawano}}, \bibinfo {author} {\bibfnamefont {Masashi}\ \bibnamefont
  {Kurosawa}}, \ and\ \bibinfo {author} {\bibfnamefont {Masahiro}\ \bibnamefont
  {Nomura}},\ }\bibfield  {title} {\enquote {\bibinfo {title} {Semiballistic
  thermal conduction in polycrystalline sige nanowires},}\ }\href {\doibase
  10.1063/1.5130659} {\bibfield  {journal} {\bibinfo  {journal} {Applied
  Physics Letters}\ }\textbf {\bibinfo {volume} {115}},\ \bibinfo {pages}
  {253101} (\bibinfo {year} {2019})},\ \Eprint
  {http://arxiv.org/abs/https://pubs.aip.org/aip/apl/article-pdf/doi/10.1063/1.5130659/14529340/253101\_1\_online.pdf}
  {https://pubs.aip.org/aip/apl/article-pdf/doi/10.1063/1.5130659/14529340/253101\_1\_online.pdf}
  \BibitemShut {NoStop}%
\bibitem [{\citenamefont {Anufriev}\ \emph {et~al.}(2024)\citenamefont
  {Anufriev}, \citenamefont {Wu}, \citenamefont {Volz},\ and\ \citenamefont
  {Nomura}}]{2024IncrThCondSiCarbide}%
  \BibitemOpen
  \bibfield  {author} {\bibinfo {author} {\bibfnamefont {Roman}\ \bibnamefont
  {Anufriev}}, \bibinfo {author} {\bibfnamefont {Yunhui}\ \bibnamefont {Wu}},
  \bibinfo {author} {\bibfnamefont {Sebastian}\ \bibnamefont {Volz}}, \ and\
  \bibinfo {author} {\bibfnamefont {Masahiro}\ \bibnamefont {Nomura}},\
  }\bibfield  {title} {\enquote {\bibinfo {title} {Quasi-ballistic thermal
  transport in silicon carbide nanowires},}\ }\href {\doibase
  10.1063/5.0180685} {\bibfield  {journal} {\bibinfo  {journal} {Applied
  Physics Letters}\ }\textbf {\bibinfo {volume} {124}},\ \bibinfo {pages}
  {022202} (\bibinfo {year} {2024})},\ \Eprint
  {http://arxiv.org/abs/https://pubs.aip.org/aip/apl/article-pdf/doi/10.1063/5.0180685/18293217/022202\_1\_5.0180685.pdf}
  {https://pubs.aip.org/aip/apl/article-pdf/doi/10.1063/5.0180685/18293217/022202\_1\_5.0180685.pdf}
  \BibitemShut {NoStop}%
\bibitem [{\citenamefont {Landau}\ and\ \citenamefont
  {Lifshitz}(1958)}]{StatPhysLandau}%
  \BibitemOpen
  \bibfield  {author} {\bibinfo {author} {\bibfnamefont {L.~D.}\ \bibnamefont
  {Landau}}\ and\ \bibinfo {author} {\bibfnamefont {E.~M.}\ \bibnamefont
  {Lifshitz}},\ }\href@noop {} {\emph {\bibinfo {title} {Statistical
  Physics}}}\ (\bibinfo  {publisher} {Reading, MA: Addison-Wesley},\ \bibinfo
  {year} {1958})\BibitemShut {NoStop}%
\bibitem [{\citenamefont {Landau}\ \emph
  {et~al.}(1986{\natexlab{b}})\citenamefont {Landau}, \citenamefont {Lifshitz},
  \citenamefont {Kosevich}, \citenamefont {Sykes}, \citenamefont {Pitaevskii},\
  and\ \citenamefont {Reid}}]{landau1986Elasticitytheory}%
  \BibitemOpen
  \bibfield  {author} {\bibinfo {author} {\bibfnamefont {L.D.}\ \bibnamefont
  {Landau}}, \bibinfo {author} {\bibfnamefont {E.M.}\ \bibnamefont {Lifshitz}},
  \bibinfo {author} {\bibfnamefont {A.M.}\ \bibnamefont {Kosevich}}, \bibinfo
  {author} {\bibfnamefont {J.B.}\ \bibnamefont {Sykes}}, \bibinfo {author}
  {\bibfnamefont {L.P.}\ \bibnamefont {Pitaevskii}}, \ and\ \bibinfo {author}
  {\bibfnamefont {W.H.}\ \bibnamefont {Reid}},\ }\href
  {https://books.google.com/books?id=tpY-VkwCkAIC} {\emph {\bibinfo {title}
  {Theory of Elasticity: Volume 7}}},\ Course of theoretical physics\ (\bibinfo
   {publisher} {Elsevier Science},\ \bibinfo {year} {1986})\BibitemShut
  {NoStop}%
\bibitem [{\citenamefont {Kagan}\ and\ \citenamefont
  {Iosilevskii}(1964)}]{KaganIoselevskiiQuasiLoc}%
  \BibitemOpen
  \bibfield  {author} {\bibinfo {author} {\bibfnamefont {Yu.}\ \bibnamefont
  {Kagan}}\ and\ \bibinfo {author} {\bibfnamefont {Ya.}\ \bibnamefont
  {Iosilevskii}},\ }\bibfield  {title} {\enquote {\bibinfo {title} {Anomalous
  behavior of specific heat of crystals with heavy impurity atoms},}\ }\href
  {jetp.ras.ru/cgi-bin/dn/e_018_02_0562.pdf} {\bibfield  {journal} {\bibinfo
  {journal} {JETP}\ }\textbf {\bibinfo {volume} {18}},\ \bibinfo {pages} {562}
  (\bibinfo {year} {1964})}\BibitemShut {NoStop}%
\bibitem [{\citenamefont {Mathews}\ \emph {et~al.}(1965)\citenamefont
  {Mathews}, \citenamefont {Seshadri},\ and\ \citenamefont
  {Vijayalakshmi}}]{1965QuasilocSignofNextNeighb}%
  \BibitemOpen
  \bibfield  {author} {\bibinfo {author} {\bibfnamefont {P.~M.}\ \bibnamefont
  {Mathews}}, \bibinfo {author} {\bibfnamefont {M.~S.}\ \bibnamefont
  {Seshadri}}, \ and\ \bibinfo {author} {\bibfnamefont {R.}~\bibnamefont
  {Vijayalakshmi}},\ }\bibfield  {title} {\enquote {\bibinfo {title} {The
  effect of second neighbour interactions on the dynamics of a linear chain
  with impurities},}\ }\href {\doibase
  https://doi.org/10.1002/zamm.19650450407} {\bibfield  {journal} {\bibinfo
  {journal} {ZAMM - Journal of Applied Mathematics and Mechanics / Zeitschrift
  für Angewandte Mathematik und Mechanik}\ }\textbf {\bibinfo {volume} {45}},\
  \bibinfo {pages} {217--223} (\bibinfo {year} {1965})},\ \Eprint
  {http://arxiv.org/abs/https://onlinelibrary.wiley.com/doi/pdf/10.1002/zamm.19650450407}
  {https://onlinelibrary.wiley.com/doi/pdf/10.1002/zamm.19650450407}
  \BibitemShut {NoStop}%
\bibitem [{\citenamefont {Polishchuk}\ \emph {et~al.}(1988)\citenamefont
  {Polishchuk}, \citenamefont {Zhernov},\ and\ \citenamefont
  {Maksimov}}]{1988IYPVibrLoc8}%
  \BibitemOpen
  \bibfield  {author} {\bibinfo {author} {\bibfnamefont {I.~Ya.}\ \bibnamefont
  {Polishchuk}}, \bibinfo {author} {\bibfnamefont {A.~P.}\ \bibnamefont
  {Zhernov}}, \ and\ \bibinfo {author} {\bibfnamefont {L.~A.}\ \bibnamefont
  {Maksimov}},\ }\bibfield  {title} {\enquote {\bibinfo {title} {Thermal
  conductivity and interference effects in the dynamics of crystals woith heavy
  impurities},}\ }\href
  {http://jetp.ras.ru/cgi-bin/e/index/e/67/9/p1881?a=list} {\bibfield
  {journal} {\bibinfo  {journal} {JETP}\ }\textbf {\bibinfo {volume} {67}},\
  \bibinfo {pages} {362} (\bibinfo {year} {1988})}\BibitemShut {NoStop}%
\bibitem [{\citenamefont {Polishchuk}\ \emph {et~al.}(1990)\citenamefont
  {Polishchuk}, \citenamefont {Burin},\ and\ \citenamefont
  {Maksimov}}]{ab90JETPLettPhonons}%
  \BibitemOpen
  \bibfield  {author} {\bibinfo {author} {\bibfnamefont {I.~Ya.}\ \bibnamefont
  {Polishchuk}}, \bibinfo {author} {\bibfnamefont {A.~L.}\ \bibnamefont
  {Burin}}, \ and\ \bibinfo {author} {\bibfnamefont {L.~A.}\ \bibnamefont
  {Maksimov}},\ }\bibfield  {title} {\enquote {\bibinfo {title} {Anderson
  localization in crystals with heavy isotopic impurities},}\ }\href
  {jetpletters.ru/ps/1172/article_17706.pdf} {\bibfield  {journal} {\bibinfo
  {journal} {JETP Letters}\ }\textbf {\bibinfo {volume} {51}},\ \bibinfo
  {pages} {730--734} (\bibinfo {year} {1990})}\BibitemShut {NoStop}%
\bibitem [{\citenamefont {Polishchuk}\ \emph
  {et~al.}(1997{\natexlab{a}})\citenamefont {Polishchuk}, \citenamefont
  {Maksimov},\ and\ \citenamefont {Burin}}]{abiypphysrep}%
  \BibitemOpen
  \bibfield  {author} {\bibinfo {author} {\bibfnamefont {I.Ya.}\ \bibnamefont
  {Polishchuk}}, \bibinfo {author} {\bibfnamefont {L.A.}\ \bibnamefont
  {Maksimov}}, \ and\ \bibinfo {author} {\bibfnamefont {A.L.}\ \bibnamefont
  {Burin}},\ }\bibfield  {title} {\enquote {\bibinfo {title} {Localization and
  propagation of phonons in crystals with heavy impurities},}\ }\href {\doibase
  https://doi.org/10.1016/S0370-1573(97)00025-2} {\bibfield  {journal}
  {\bibinfo  {journal} {Physics Reports}\ }\textbf {\bibinfo {volume} {288}},\
  \bibinfo {pages} {205 -- 222} (\bibinfo {year}
  {1997}{\natexlab{a}})}\BibitemShut {NoStop}%
\bibitem [{\citenamefont {Bernardin}\ \emph {et~al.}(2019)\citenamefont
  {Bernardin}, \citenamefont {Huveneers},\ and\ \citenamefont
  {Olla}}]{2019AnalyticsForRandMass}%
  \BibitemOpen
  \bibfield  {author} {\bibinfo {author} {\bibfnamefont {C{\'e}dric}\
  \bibnamefont {Bernardin}}, \bibinfo {author} {\bibfnamefont {Fran{\c{c}}ois}\
  \bibnamefont {Huveneers}}, \ and\ \bibinfo {author} {\bibfnamefont {Stefano}\
  \bibnamefont {Olla}},\ }\bibfield  {title} {\enquote {\bibinfo {title}
  {Hydrodynamic limit for a disordered harmonic chain},}\ }\href {\doibase
  10.1007/s00220-018-3251-4} {\bibfield  {journal} {\bibinfo  {journal}
  {Communications in Mathematical Physics}\ }\textbf {\bibinfo {volume}
  {365}},\ \bibinfo {pages} {215--237} (\bibinfo {year} {2019})}\BibitemShut
  {NoStop}%
\bibitem [{\citenamefont {Shishkina}\ and\ \citenamefont
  {Gavrilov}(2023)}]{2023QuasilocModIsotopDefect}%
  \BibitemOpen
  \bibfield  {author} {\bibinfo {author} {\bibfnamefont {Ekaterina~V.}\
  \bibnamefont {Shishkina}}\ and\ \bibinfo {author} {\bibfnamefont {Serge~N.}\
  \bibnamefont {Gavrilov}},\ }\bibfield  {title} {\enquote {\bibinfo {title}
  {Unsteady ballistic heat transport in a 1d harmonic crystal due to a source
  on an isotopic defect},}\ }\href {\doibase 10.1007/s00161-023-01188-x}
  {\bibfield  {journal} {\bibinfo  {journal} {Continuum Mechanics and
  Thermodynamics}\ }\textbf {\bibinfo {volume} {35}},\ \bibinfo {pages}
  {431--456} (\bibinfo {year} {2023})}\BibitemShut {NoStop}%
\bibitem [{\citenamefont {Abrahams}\ \emph {et~al.}(1979)\citenamefont
  {Abrahams}, \citenamefont {Anderson}, \citenamefont {Licciardello},\ and\
  \citenamefont {Ramakrishnan}}]{Abrahams1979ScThLoc}%
  \BibitemOpen
  \bibfield  {author} {\bibinfo {author} {\bibfnamefont {E.}~\bibnamefont
  {Abrahams}}, \bibinfo {author} {\bibfnamefont {P.~W.}\ \bibnamefont
  {Anderson}}, \bibinfo {author} {\bibfnamefont {D.~C.}\ \bibnamefont
  {Licciardello}}, \ and\ \bibinfo {author} {\bibfnamefont {T.~V.}\
  \bibnamefont {Ramakrishnan}},\ }\bibfield  {title} {\enquote {\bibinfo
  {title} {Scaling theory of localization: Absence of quantum diffusion in two
  dimensions},}\ }\href {\doibase 10.1103/PhysRevLett.42.673} {\bibfield
  {journal} {\bibinfo  {journal} {Phys. Rev. Lett.}\ }\textbf {\bibinfo
  {volume} {42}},\ \bibinfo {pages} {673--676} (\bibinfo {year}
  {1979})}\BibitemShut {NoStop}%
\bibitem [{\citenamefont {Gor'kov}\ \emph {et~al.}(1979)\citenamefont
  {Gor'kov}, \citenamefont {Larkin},\ and\ \citenamefont
  {Khmel'nitskii}}]{Gorkov792D}%
  \BibitemOpen
  \bibfield  {author} {\bibinfo {author} {\bibfnamefont {L.~P.}\ \bibnamefont
  {Gor'kov}}, \bibinfo {author} {\bibfnamefont {A.~I.}\ \bibnamefont {Larkin}},
  \ and\ \bibinfo {author} {\bibfnamefont {D.~E.}\ \bibnamefont
  {Khmel'nitskii}},\ }\bibfield  {title} {\enquote {\bibinfo {title} {Particle
  conductivity in a two-dimensional random potential},}\ }\href
  {http://www.jetpletters.ac.ru/ps/1364/article_20629.pdf} {\bibfield
  {journal} {\bibinfo  {journal} {JETP Letters}\ }\textbf {\bibinfo {volume}
  {30}},\ \bibinfo {pages} {228} (\bibinfo {year} {1979})}\BibitemShut
  {NoStop}%
\bibitem [{\citenamefont {Flach}\ \emph {et~al.}(2003)\citenamefont {Flach},
  \citenamefont {Miroshnichenko},\ and\ \citenamefont
  {Fistul}}]{2003DiscreteBreathFlach}%
  \BibitemOpen
  \bibfield  {author} {\bibinfo {author} {\bibfnamefont {S.}~\bibnamefont
  {Flach}}, \bibinfo {author} {\bibfnamefont {A.~E.}\ \bibnamefont
  {Miroshnichenko}}, \ and\ \bibinfo {author} {\bibfnamefont {M.~V.}\
  \bibnamefont {Fistul}},\ }\bibfield  {title} {\enquote {\bibinfo {title}
  {Wave scattering by discrete breathers},}\ }\href {\doibase
  10.1063/1.1561627} {\bibfield  {journal} {\bibinfo  {journal} {Chaos: An
  Interdisciplinary Journal of Nonlinear Science}\ }\textbf {\bibinfo {volume}
  {13}},\ \bibinfo {pages} {596--609} (\bibinfo {year} {2003})},\ \Eprint
  {http://arxiv.org/abs/https://pubs.aip.org/aip/cha/article-pdf/13/2/596/18302698/596\_1\_online.pdf}
  {https://pubs.aip.org/aip/cha/article-pdf/13/2/596/18302698/596\_1\_online.pdf}
  \BibitemShut {NoStop}%
\bibitem [{\citenamefont {Polishchuk}\ \emph
  {et~al.}(1997{\natexlab{b}})\citenamefont {Polishchuk}, \citenamefont
  {Maksimov},\ and\ \citenamefont {Burin}}]{ab97PhTrPhysRep}%
  \BibitemOpen
  \bibfield  {author} {\bibinfo {author} {\bibfnamefont {I.Ya.}\ \bibnamefont
  {Polishchuk}}, \bibinfo {author} {\bibfnamefont {L.A.}\ \bibnamefont
  {Maksimov}}, \ and\ \bibinfo {author} {\bibfnamefont {A.L.}\ \bibnamefont
  {Burin}},\ }\bibfield  {title} {\enquote {\bibinfo {title} {Localization and
  propagation of phonons in crystals with heavy impurities},}\ }\href {\doibase
  https://doi.org/10.1016/S0370-1573(97)00025-2} {\bibfield  {journal}
  {\bibinfo  {journal} {Physics Reports}\ }\textbf {\bibinfo {volume} {288}},\
  \bibinfo {pages} {205--222} (\bibinfo {year} {1997}{\natexlab{b}})},\
  \bibinfo {note} {i.M. Lifshitz and Condensed Matter Theory}\BibitemShut
  {NoStop}%
\bibitem [{\citenamefont {Lee}(1970)}]{1970LEE1DModelofVibrWithnnextint}%
  \BibitemOpen
  \bibfield  {author} {\bibinfo {author} {\bibfnamefont {S.M.}\ \bibnamefont
  {Lee}},\ }\bibfield  {title} {\enquote {\bibinfo {title} {Normal vibrations
  of a finite linear lattice with nearest- and next-nearest-neighbor
  interactions},}\ }\href {\doibase
  https://doi.org/10.1016/0022-460X(70)90084-2} {\bibfield  {journal} {\bibinfo
   {journal} {Journal of Sound and Vibration}\ }\textbf {\bibinfo {volume}
  {12}},\ \bibinfo {pages} {143--151} (\bibinfo {year} {1970})}\BibitemShut
  {NoStop}%
\bibitem [{\citenamefont {Lee}(1976)}]{19762DversionLEE}%
  \BibitemOpen
  \bibfield  {author} {\bibinfo {author} {\bibfnamefont {S.M.}\ \bibnamefont
  {Lee}},\ }\bibfield  {title} {\enquote {\bibinfo {title} {Normal vibration
  frequencies of a rectangular two-dimensional array of identical
  point-masses},}\ }\href {\doibase
  https://doi.org/10.1016/0022-460X(76)90738-0} {\bibfield  {journal} {\bibinfo
   {journal} {Journal of Sound and Vibration}\ }\textbf {\bibinfo {volume}
  {45}},\ \bibinfo {pages} {595--600} (\bibinfo {year} {1976})}\BibitemShut
  {NoStop}%
\bibitem [{\citenamefont {Dwivedi}\ and\ \citenamefont
  {Chua}(2016)}]{2016GeneralizedTransfMatr1D}%
  \BibitemOpen
  \bibfield  {author} {\bibinfo {author} {\bibfnamefont {Vatsal}\ \bibnamefont
  {Dwivedi}}\ and\ \bibinfo {author} {\bibfnamefont {Victor}\ \bibnamefont
  {Chua}},\ }\bibfield  {title} {\enquote {\bibinfo {title} {Of bulk and
  boundaries: Generalized transfer matrices for tight-binding models},}\ }\href
  {\doibase 10.1103/PhysRevB.93.134304} {\bibfield  {journal} {\bibinfo
  {journal} {Phys. Rev. B}\ }\textbf {\bibinfo {volume} {93}},\ \bibinfo
  {pages} {134304} (\bibinfo {year} {2016})}\BibitemShut {NoStop}%
\bibitem [{\citenamefont {Goldstein}\ \emph {et~al.}(2015)\citenamefont
  {Goldstein}, \citenamefont {Huse}, \citenamefont {Lebowitz},\ and\
  \citenamefont {Tumulka}}]{Lebowitz15}%
  \BibitemOpen
  \bibfield  {author} {\bibinfo {author} {\bibfnamefont {Sheldon}\ \bibnamefont
  {Goldstein}}, \bibinfo {author} {\bibfnamefont {David~A.}\ \bibnamefont
  {Huse}}, \bibinfo {author} {\bibfnamefont {Joel~L.}\ \bibnamefont
  {Lebowitz}}, \ and\ \bibinfo {author} {\bibfnamefont {Roderich}\ \bibnamefont
  {Tumulka}},\ }\bibfield  {title} {\enquote {\bibinfo {title} {Thermal
  equilibrium of a macroscopic quantum system in a pure state},}\ }\href
  {\doibase 10.1103/PhysRevLett.115.100402} {\bibfield  {journal} {\bibinfo
  {journal} {Phys. Rev. Lett.}\ }\textbf {\bibinfo {volume} {115}},\ \bibinfo
  {pages} {100402} (\bibinfo {year} {2015})}\BibitemShut {NoStop}%
\end{thebibliography}%
\end{document}